\documentclass{emulateapj}

\newcommand{\gfvalues}{$gf$-values}
\newcommand{\loggf}{log $gf$-values}
\newcommand{\etal}{et al.}

\newcommand\teff{T$_{\rm eff}$}
\newcommand\logg{log $g$}
\newcommand\afe{[$\alpha$/Fe]}
\newcommand{\mfeh}{\rm{[m/H]}}

\newcommand{\mlogg}{\rm{log}\thinspace g}
\newcommand{\wcen}{$\omega$\thinspace Cen}

\shorttitle{Chemical Compositions of Red Giants in Old LMC GCs}
\shortauthors{Johnson et al.}

\begin{document}
\title{Chemical Compositions of Red Giant Stars in \hfill \\
Old Large Magellanic Cloud Globular Clusters}

\author{{Jennifer A.\ Johnson}
\altaffilmark{1}}
\altaffiltext{1}{Present Address: Department of Astronomy, Ohio State University, 140 West 18th Avenue, Columbus, OH 43210}
\affil{Dominion Astrophysical Observatory, Herzberg Institute of Astrophysics, National Research Council, 5071 West Saanich Rd., Victoria, BC V9E 2E7, Canada}
\affil{Observatories of the Carnegie Institute of Washington,
  Pasadena, CA 91101 USA }
\email{Jennifer.Johnson@nrc-cnrc.gc.ca}

\author{{Inese I.\ Ivans}
\altaffilmark{2,3,4}}
\altaffiltext{2}{Hubble Fellow}
\altaffiltext{3}{Present Address: The Observatories of the the Carnegie
Instution of Washington, Pasadena, CA 91101}
\altaffiltext{4}{Present Address: Princeton University 
Observatory, Peyton Hall, Princeton, NJ 08544}
\affil{Department of Astronomy, California Institute of Technology, Pasadena, CA 91125 USA}
\email{iii@ociw.edu}

\and

\author{Peter B.\ Stetson}
\affil{Dominion Astrophysical Observatory, Herzberg Institute of Astrophysics, National Research Council, 5071 West Saanich Rd., Victoria, BC V9E 2E7, Canada}
\email{Peter.Stetson@nrc-cnrc.gc.ca}

\begin{abstract}
We have observed ten red giant stars in four old Large Magellanic
Cloud globular clusters with the high-resolution spectrograph MIKE
on the Magellan Landon Clay 6.5-m telescope. The stars in our sample have up to 20 elemental abundance determinations
for the $\alpha$-, iron-peak, and neutron-capture element groups.  We have also derived abundances for the light odd-Z elements Na and Al.  We find NGC 2005 and NGC 2019 to be more metal-rich than previous estimates from the Ca\thinspace{\sc ii} triplet, and we derive [Fe/H] values closer to those obtained 
from the slope of the red giant branch.  However, we confirm previous determinations for Hodge 11 and NGC 1898 to within 0.2~dex.  
The LMC cluster 
\lbrack Mg/Fe\rbrack{} and \lbrack Si/Fe\rbrack{} ratios are comparable
to the values observed in old Galactic globular
cluster stars, as are the abundances [Y/Fe], [Ba/Fe], and [Eu/Fe].  The LMC clusters do not share the low-Y behavior observed in some dwarf spheroidal galaxies. \lbrack Ca/Fe\rbrack, \lbrack Ti/Fe\rbrack, and 
\lbrack V/Fe\rbrack{} in the LMC, however, are {\em significantly} lower than what is seen in the
Galactic globular cluster system.
Neither does the behavior of \lbrack Cu/Fe\rbrack{} as a function of
[Fe/H] in our LMC clusters match the
trend seen in the Galaxy, staying instead at a constant value of 
$\sim$$-$0.8.
Because not all \lbrack$\alpha$/Fe\rbrack{} ratios are suppressed, 
these abundance ratios cannot be attributed solely
to the injection of Type Ia SNe material, and instead reflect the
differences in  star formation history of the LMC vs.\ the Milky Way.
An extensive numerical experimental study was performed, varying both input parameters and
stellar atmosphere models, to verify that the unusual abundance ratios 
derived in this study are not the result of the adopted 
atomic parameters, stellar atmospheres or stellar parameters.
We conclude that many of the abundances in the LMC globular clusters we observed are distinct from 
those observed in the Milky Way, and these differences are intrinsic 
to the stars in those systems. 
\end{abstract}

\section{Introduction}

Globular clusters have been key to gaining insights into the early
epoch of formation and evolution for galaxies in general and for the 
Galaxy in particular.  
Because of the proximity of Galactic globular clusters (GGC) 
to us, we can obtain color-magnitude
diagrams and high-resolution spectra of individual stars, which has
allowed us to measure ages and abundances with unique accuracy. These
data show a complex and interesting picture for the GGC, including
a dispersion in abundance ratios, trends in ratios with kinematics,
and the possibility of the capture of clusters from other
galaxies. We can now observe clusters in other galaxies of the Local Group 
with the same
techniques to compare their cluster systems with the GGC and determine
the variation in globular cluster systems from galaxy to galaxy and
the possible contributions of other galaxies to the Milky Way system.

The Magellanic Clouds, less distant than some GGCs,
provide an excellent opportunity to observe abundance patterns
in another globular cluster system in detail.
The Large Magellanic Cloud (LMC) 
has long been known 
to harbour clusters of similar age, mass and metallicity 
to the GGCs (Searle, Wilkinson,
\& Bagnuolo 1980).  Testa \etal\ (1995) and 
Brocato \etal\ (1996) provided the first ages based on  
main-sequence turnoff
measurements of the oldest clusters in the LMC.  The main-sequence 
turnoffs in a large number of old clusters in the LMC have
subsequently been observed with the Hubble Space Telescope (HST).  
Some clusters in the LMC are coeval with nearby GGCs,
such as M5, M4, and M92 (Olsen \etal{} 1998, LMC-O98, hereafter; 
Johnson \etal{} 1999, LMC-J99, hereafter).

The ages, kinematics, metallicities and abundance ratios of the GGCs
have provided much insight into the formation of the Galaxy.
Searle \& Zinn (1978) argued that
the outer halo clusters were younger than the
inner halo clusters and that implied that a slow, chaotic buildup of the
outer parts of the Galaxy had occurred. That mergers have contributed to the
formation of the Galaxy was clearly shown with the discovery that the
Sagittarius dwarf spheroidal galaxy (dSph) is currently
being subsumed by the Milky Way (Ibata \etal\ 1994). 
The positions of GGCs 
on great circle orbits (Buonanno \etal{}
1994) that sometimes include other satellites of the Milky Way (Fusi Pecci \etal\ 1995) hint
at past accretion events. Lin \& Richer (1992) argued that the positions
and radial velocities of Rup 106 and possibly Pal 12 suggest that they
had been tidally captured from the Magellanic Clouds (MC). By incorporating
proper motions in the analysis, 
Dinescu \etal\ (2000) suggested that it was more
likely that Sagittarius was the original host galaxy of Pal 12. Bellazzini,
Ferraro, \& Ibata (2003) extended the analysis to conclude that at least
four outer halo GGCs belonged to Sagittarius, in addition to the 
four clusters whose positions lie near the main body of Sagittarius 
(Ibata \etal\ 1995) 

Information about the history of the Galaxy is also contained in the
chemical abundance ratios of old stars.  In a seminal paper, Tinsley
(1979) argued that enhanced \afe{}-ratios{\footnote{We adopt the usual
spectroscopic notation that [A/B] $\equiv$ {\rm log}$_{\rm
10}$(N$_{\rm A}$/N$_{\rm B}$)$_{\star}$ -- {\rm log}$_{\rm
10}$(N$_{\rm A}$/N$_{\rm B}$)$_{\odot}$, and log~$\epsilon$(A)
$\equiv$ {\rm log}$_{\rm 10}$(N$_{\rm A}$/N$_{\rm H}$) + 12.0, for
elements A and B.  Also, in this paper, except for instances where 
[m/H] or Z are specifically stated, we define metallicity as the 
stellar [Fe/H].}  in metal-poor stars were a consequence of the
different timescales for the production of the $\alpha$-elements
(e.g., O, Ne, Mg, Si, Ca, and sometimes Ti) in core-collapse
supernovae (Type II SNe) vs.\ the Fe produced by both SNe Type Ia and
Type II.  SNe Type II progenitors are short-lived (1Myr--100Myr) massive
stars whereas progenitors of SNe Type Ia (mass-exchange binary systems
including a white dwarf star) require longer to evolve and do not
contribute to the chemical evolution of the Galaxy until $\ge 1$Gyr
subsequent to the formation of the binary system (Timmes, Woosley, \&
Weaver 1995; Matteucci \& Recchi 2001).  Hence, the ratio of Type
Ia/Type II SNe events determines the \afe{}.  Systems that have
recently started forming stars and have only had the contributions
from massive stars to the interstellar medium would then be predicted
to possess low Type Ia/II ratios and \afe{}\ $>0$. Old GCCs possess
high \afe{} ratios (Pilachowski, Sneden, \& Wallerstein 1983; and
references therein), in accord with the idea that GGCs were among the
first surviving Galactic objects to have formed.

Nissen
\& Schuster (1997; hereafter, NS97) discovered a sample of moderately metal-poor
 field stars with low [$\alpha$/Fe] ratios and
suggested that they could have accreted from
dwarf galaxies with a chemical
evolution history different than that of the solar neighborhood, allowing the
material out of which they formed
to include the ejecta from Type Ia SNe while they were still
relatively metal-poor. Also exhibiting low \afe{} ratios
with respect to the general GGC population are Rup 106 and Pal 12
(Brown, Wallerstein, \& Zucker 1997), 
which are 2-3 Gyr younger than other GGCs (Buonanno \etal\ 1990;
Stetson
\etal\ 1989). 
Brown \etal\ interpreted the solar 
\afe{}-ratios to be the result of the cluster being formed long
enough after star formation had begun in the
surrounding region to have its abundance ratios substantially
affected by contributions of iron from Type Ia SNe.

The characteristics of the production sites of other elements may also
provide insight into timescales, initial mass functions, and other 
properties of clusters. At this time, both the predictions from theory
and the data from globular clusters are murkier than the case of
[$\alpha$/Fe]. For example, the site of the rapid neutron-capture
process ($r$-process) is uncertain, but it is clear that the 
r-processed material
appears before the slow-neutron-capture process ($s$-process) 
in asymptotic giant
branch stars begins to contribute much to Galactic chemical 
evolution (Truran 1981). The r-process produces some heavy elements,
such as Eu, more readily than others, such as Ba and La, while
the s-process does the opposite. Therefore, ratios such as [Ba/Fe] and
[Ba/Eu] contain information about when clusters formed from
a chemical evolution standpoint. Other element ratios are also 
sensitive to the mix of stars that polluted the ISM.
First, the metallicity of the progenitor of a Type II SNe 
is important in the synthesis
of such elements as Na, Al, and Cu (e.g., Arnett 1971, Woosley \&
Weaver 1995). Second, the mass of the SN affects the ratio of the
$\alpha$ elements produced (e.g. Woosley \& Weaver 1995; McWilliam
1997). Less massive stars make lower ratios of [Mg/Ca] and [Mg/Si],
for example. [Si/Ti] should be highest for a 20 M$_{\odot}$ star
according to the Woosley \& Weaver yields.

Recent efforts to measure many elements in globular clusters
have shown that other abundance ratios, such as the ones listed
 above, vary between GGCs.  
Ivans et al.\ (1999; 2001; hereafter M4-I99 and M5-I01), measured 
abundance ratios of 14 elements in 36 giants in each of M4 and M5, 
two GGCs with similar [Fe/H] and ages. Within either cluster, 
the stars possess comparable abundance ratios for elements
not sensitive to proton-capture nucleosynthesis, but the same is not true of a 
comparison between clusters. 
Confirming and expanding upon the earlier results of Brown \& 
Wallerstein (1992), M4-I99 found that the mean [Si/Fe] ratio for the 
M4 stars is 3-$\sigma$ greater than in M5 stars.  The abundances of [Al/Fe], 
[Ba/Fe], and [La/Fe] are also significantly higher in M4 stars. 
Interestingly, these clusters also differ in their apogalactic 
distances, with apogalactocentric radii of 5.9 and 35.4 kpc for M4 
and M5, respectively (Dinescu et al. 1999).  
This same apparent trend with apogalactic distance and some abundance ratios
may also be reflected in halo field and cluster stars 
(NS97; Hanson \etal\
1998; Stephens 1999; Fulbright 2002; Lee \& Carney
2002; Fulbright 2004).  However, employing an extensive sample from
the literature which included the results incorporated in Stephens
(1999) and Fulbright (2004), Venn \etal{} (2004) argue
that the low [$\alpha$/Fe] ratios 
in halo stars are, if anything, correlated with extreme 
retrograde orbits and statistically not correlated with apogalactic distance.

Additional cluster-to-cluster abundance variations have been found in other studies.  Pursuing the investigation of the \afe\ trends to the inner halo, Lee 
\& Carney (2002) report high [Si/Fe] and low [Ti/Fe] in NGC 6287, NGC6293, and NGC6541, three 
metal-poor GCC (--1.8 $\le$ [Fe/H] $\le$ --2.0). Lee, Carney, \& 
Habgood (2004) report high [Si/Fe] and low [Ti/Fe] in M68 stars as well.  
Pal 12 stars, in addition to low \afe{} compared with other GGCs, also 
possess subsolar values of [Na/Fe] and [Ni/Fe] (Brown et al. 1997; 
Cohen 2004). Ter 7 stars show low [Ni/Fe] (Tautvai{\v s}ien{\.e} 
et al.\ 2004).  As these examples illustrate, the evidence for abundance variations between GGCs has 
been firmly established

Within an individual GGC, star-to-star variations are observed among
the light elements sensitive to proton-capture nucleosynthesis (e.g., 
C, N, O, Na, Mg, and Al).  Early detections of CN variations among
giant stars by Lindblad (1922) and Popper (1947) were expanded to
higher resolution  (see e.g., Osborn 1971; 
Peterson 1980) and, for some GGCs, to stars on the main sequence (see 
Hesser 1978; Hesser \& Bell 1980).  Also observed in 
globular cluster populations (but absent in the field star population) 
are anti-correlations in the abundances of [O/Fe] with [Na/Fe] and 
[Al/Fe] (see Gratton, Sneden, \& Carretta 2004 for a recent review).

One possible explanation for the light-element abundance patterns is
``deep'' mixing in red giants (e.g., Sweigart \& Mengel 1979;
Denissenkov \& Weiss 1996, Denissenkov \& VandenBerg 2003), dredging
up the products of proton-capture nucleosynthesis from the interior
out to the photosphere.  It remains unclear, however, how the temperatures
of the interiors of the red giant stars can even get hot enough to
convert Mg to Al (Langer, Hoffman, \& Zaidins 1997; Messenger \&
Lattanzio 2002).  For some time, it had been thought that
intermediate-mass asymptotic giant branch (AGB) stars could be
responsible for producing the abundance patterns (Cottrell \& Da Costa
1981), but recent work by Denissenkov \& Herwig (2003) and Fenner
\etal\ (2004) show that the observed abundance correlations are not
replicated in model yields of AGB stars.
 
The presence of the abundance correlations at or below the 
main-sequence turnoff suggests that the variations may be primordial 
or the result of pollution by more evolved stars.  It is likely that 
some combination of effects are at work.  In M13, for example, the 
abundance patterns are also correlated with the evolutionary state of the 
stars (Kraft \etal\ 1993; Sneden \etal\ 2004; Johnson \etal\ 2005). 
While all clusters
that have been examined for deep mixing effects show the associated abundance
anomalies, some clusters appear to be more affected
than others. The classic example is M3 and M13 (Kraft \etal\ 1992),
where stars in M13 have [O/Fe] down to values of $-$0.87, while the
most oxygen-poor stars M3 stars have [O/Fe] = $-0.25$. Other abundance ratios
of light elements sensitive to proton-capture nucleosynthesis are similarly extreme in M13 but not in M3 (see e.g., Johnson \etal\ 2005 and references
therein).

To summarize the situation in the GCC system, the majority of clusters
exhibit super-solar \afe{} ratios, abundance ratio
trends with apogalactic distance or prograde/retrograde orbits, 
solar iron-peak element ratios, 
and evidence of abundance correlations
in the light elements, possibly due to deep mixing.
{\it Are these universal properties of old 
globular cluster systems, or do they represent the unique history of 
the Milky Way?}

Low-dispersion spectra of individual giants by Cowley \& Hartwick 
(1982)
 were used to measure spectral indices for nine old LMC clusters, 
including Hodge 11.  Subsequently, Olszewski \etal\ (1991, LMC-O91) 
performed a comprehensive study to measure the metallicities of the LMC 
clusters using low-dispersion spectra of the \ion{Ca}{2} infrared triplet lines in individual giants.  These measurements have been extremely 
useful in tracing the age-metallicity relationship in the LMC and in 
providing estimates of the overall metallicity.  However, for some
Magellanic Cloud clusters (e.g., NGC~2019 and NGC~2005), there is a disagreement 
between the metallicities from LMC-O91 and the slopes of the red 
giant branches measured by LMC-O98.  Abundance ratio 
questions could not be addressed, however, until high-resolution studies 
of LMC stars became available.  The study by Hill \etal\ (2000) 
included LMC clusters of a range of ages, including one old cluster: 
NGC 2210 ([Fe/H] = $-$1.75). The three stars observed in NGC 2210 have
[O/Fe] values of 0.02, 0.19 and 0.21 dex, 
lower than are typical in GGC red giant stars.  
Smith \etal\ (2002) observed one star in NGC 1898 
in the near-IR with PHOENIX on Gemini. [Ti/Fe] is low in this star, 
illustrating that, for this cluster at least, 
LMC clusters do not always exhibit the high \afe{} 
abundances that typically belong to old GGCs. 

In this paper, we report on abundances in four clusters in the Large
Magellanic Cloud (LMC) observed with the MIKE spectrograph on
Magellan. These
clusters, NGC 1898, NGC 2005, NGC 2019 and Hodge 11 are globular
clusters with ages from color-magnitude diagrams that are as old as
the majority of GGCs (LMC-O98, LMC-J99) and
have metallicities ranging from $-2.0 $ to $-1.0$. Therefore, they
are similar to the kinds of GGC that have been important in 
deciphering the history of the Galactic spheroid.

\section{Observations and Reductions}

\subsection{Target Selection}

To expand the sample of LMC cluster stars with high-resolution spectra 
and abundance ratio determinations for a large number of elements,
we observed four old LMC clusters: Hodge 11, 
NGC 1898, NGC 2005, and NGC 2019. Hodge 11 is located 4.7$^{o}$ from the center.  The other three clusters are located within 1.5$^{o}$ of the
center of the LMC and are suitable comparisons to the inner halo GCCs.

LMC clusters are far enough away that stellar crowding leading to blended
spectra is a potential problem, and this was one of the main concerns 
in the sample selection of LMC-O91. However, we cannot simply choose 
stars in the outskirts of the clusters, especially for clusters in
the inner part of the LMC, since the ratio of field stars to cluster 
stars increases rapidly with radius. 
Instead, we employed the HST images obtained 
by LMC-O98 for NGC 1898, NGC 2005, and NGC 2019 to identify the 
brightest stars in the inner clusters whose stellar profiles are not 
apparently blended with the profiles of other bright stars. 
For  
Hodge 11, well removed from the LMC bar and disk,
we chose the 
brightest uncrowded stars from a private catalog
maintained by one of the authors (Stetson).  
Both Hodge 11 
stars are saturated in the HST observations of LMC-J99.  Only one of the stars
is included in the observations of Mighell \etal\ (1996), and it is saturated
in the V exposures. 
As can be seen in Figure 1, all of our observed clusters stars lie at the tip of the giant branch.

\begin{figure}
\plotone{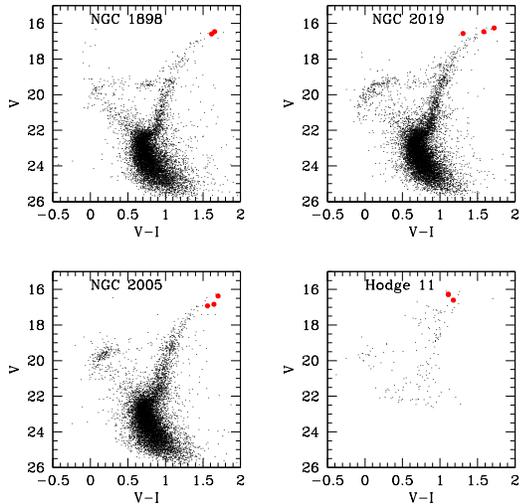}
\caption{The color-magnitude diagrams for NGC 1898, NGC 2005, NGC
  2019, and Hodge 11. The photometry is from LMC-O98 and a privately maintained catalog of one of the authors (Stetson).  The stars we observed are
indicated by large filled circles.}
\end{figure}

\subsection{Observations and Reductions}

We observed the LMC globular cluster stars 27 December
2002 -- 1 January 2003 with MIKE, the double echelle spectrograph 
(Bernstein \etal\ 2003) on the Magellan Landon Clay 6.5-m Telescope at the Las 
Campanas Observatory.  The observations are summarized in Table 1, where 
we also list other names by which the stars are known in the 
literature, taken from Lloyd-Evans (1980), Mighell 
\etal\ (1996), and LMC-O98.  

\begin{deluxetable*}{lcrlrcc}
\tablenum{1}
\tablewidth{0pt}
\tablecaption{LMC Cluster Observations}
\tablehead{
\colhead{Star} & \colhead{Other} & \colhead{V} & \colhead{Obs.} & \colhead{Exposures} & \colhead{S/N}  & \colhead{$F$} \\
\colhead{}    & \colhead{Name}  & \colhead{(mag)} & \colhead{Date} & \colhead{} & \multicolumn{2}{c}{($\sim$6600\thinspace \AA)}  
}
\startdata
NGC 1898\#1 & O8535   &16.59 & 2002 Dec 31 & 3$\times$2400s & 70 & 319 \\
NGC 1898\#2 & O8526   &16.46 & 2003 Jan 1  & 3$\times$1800s & 50 & 228 \\
NGC 2005\#1 & O9347   &16.37 & 2002 Dec 28 & 2$\times$3600s & 70 & 319 \\
NGC 2005\#2 & O9348   &16.92 & 2003 Jan 1  & 3$\times$1800s & 40 & 182 \\
NGC 2005\#3 & O9353   &16.83 & 2002 Dec 30 & 3$\times$2700s & 35 & 159 \\
NGC 2019\#1 & O9697   &16.56 & 2002 Dec 29 & 3$\times$2700s & 45 & 205 \\
NGC 2019\#2 & O9688   &16.46 & 2002 Dec 31 & 3$\times$2400s & 65 & 296 \\
NGC 2019\#3 & O9692   &16.24 & 2002 Dec 28 & 1$\times$3600s &    &     \\
             &         &      & 2002 Dec 29 & 2$\times$1800s & 75 & 341 \\
Hodge 11\#1 & Mighell &$<$16.61\tablenotemark{a}
			      & 2002 Dec 31 & 1$\times$1800s &    &     \\
            &  219234439 
                       & 16.28\tablenotemark{b}
			      & 2003 Jan 1  & 1$\times$1200s &    &     \\
            &          &      & 2003 Jan 1  & 1$\times$800s  & 45 & 205 \\
Hodge 11\#2 & LE 2    & 16.61\tablenotemark{b}    
                              & 2002 Dec 29 & 3$\times$1800s & 60 & 273 \\
\enddata 
\tablenotetext{a}{Saturated image.  See \S~2.1.}
\tablenotetext{b}{Photometry from private catalog maintained by Stetson.}
\end{deluxetable*}

The blue side of the double echelle design covers 3200--4800\thinspace
\AA, and
the red, 4500--7240\thinspace \AA, with no gaps.  For the analyses presented 
here, we only employ the relatively higher signal-to-noise (S/N $> 20$)
data redward of 4800\thinspace \AA.   We 
used a 1.0 arcsec wide slit, giving a spectral 
resolution of 19,000 ($R$ $\equiv$ $\lambda/\Delta\lambda$). 
The chip had a gain of 1.06 electrons/ADU and a read noise of 4 electrons. We binned on-chip in 2x2 pixels.  Table 1 lists the S/N per pixel
we achieved in the combined spectra at 6600\thinspace \AA, along with a measure of the quality factor per resolution element ($F$ $\equiv$ $(R/\lambda)\times($S/N$)$), also referred to as a figure-of-merit.

The data were reduced using standard IRAF\footnote{IRAF is 
distributed by NOAO, which is operated by AURA, under 
cooperative agreement with the NSF.} routines. We also employed 
IRAF to combine multiple spectra taken of the same 
objects and to excise cosmic ray features. Spectra taken of 
Th-Ar lamps provided the wavelength calibration.  We took several 
spectra of the hot, rapid rotator HR 1307 to eliminate the 
telluric features using a current version of the program 
{\sf SPECTRE} (Fitzpatrick \& Sneden 1987).  

\section{Abundance Analysis}

In this section we describe our analysis techniques.  In the
abundance determinations, we employed a combination of equivalent 
width and spectrum synthesis analyses.  Also presented here are 
the linelists we used, including hyperfine structure information 
(HFS) where applicable, and equivalent widths (EWs).  We include a brief 
discussion of the effect of different choices in the stellar 
parameters on the derived abundances, and expand upon that 
discussion in the appendix.

\subsection{Equivalent Widths}

To identify transitions suitable for
EW analyses, 
we 
synthesized the spectrum (4500--7240\thinspace \AA) of NGC 1898\#1 with a current 
version of {\sf MOOG} (Sneden 1973) and identified relatively
isolated lines stronger than $\sim$25 m\AA.
These features were then measured in each of our spectra 
using {\sf SPECTRE}.  
For most species, including \ion{Ca}{1}, \ion{Ti}{1}, and \ion{Fe}{1}, 
EW measurements were sufficient for a reliable abundance analysis.
For other elements, some or all of the lines were synthesized.  For 
these lines a pseudo-EW was calculated via the {\sf
ewfind} driver in {\sf MOOG} for inclusion in Table 2.  Table 2 lists the lines we employed 
in the abundance analyses. Pseudo-EWs are given for lines marked
``SYN''.




The uncertainty in the EW for each line can be
determined by the following relationship (corrected from Equation 
3 in Fulbright \& Johnson, 2003):
\begin{equation}
\delta EW ^2 = \delta x ^2 \left (\Sigma \delta r_i ^2 +(n^2-n+1) 
	\Sigma \delta C_i ^2 \right ),
\end{equation}
\noindent where $\delta x$ is the dispersion in \AA/pix, $C_i$ is 
the value of the continuum, and $r_i$ is the intensity at pixel $i$. 
This is summed over the $n$ pixels which contain absorption in the 
line. In practice, we summed over 2.5 $\times$ FWHM of the line, where the FWHM was
given by the gaussian fit via {\sf SPECTRE}.  

The uncertainty in the continuum placement is difficult to determine.
We employed the following procedure.  We selected a subset of Ti and 
Fe lines with oscillator strengths from Wickliffe \& Lawler (1997)
 and the papers of the Oxford group (Blackwell, Petford, \& Shallis 1979;
Blackwell \etal\ 1979; Blackwell, Petford, \& Simmons 1982; 
Blackwell \etal\ 1982a, 1982b, 1982c; Blackwell
\etal\ 1986). These oscillator strengths, especially in a 
relative sense, are well determined, with uncertainties of $<$ 0.05 dex 
(which are smaller than the uncertainties imposed by the EW 
measurements).  We varied the number of continuum pixels from 10 at 
4500\thinspace \AA{} to 25 at 7000\thinspace \AA, values chosen based on our previous 
experience with high resolution spectral syntheses.  With this algorithm, the 
average uncertainty in the abundance produced by the uncertainty in the EW is 
equal to the standard deviation of the sample of lines.  We then 
determined the expected uncertainty for each individual EW.  

Lines that were not significant at the 2-$\sigma$ level were 
eliminated.  The number of lines eliminated by this criterion varied 
from one to six per star. We also used Equation 1 to determine 
limits for certain species where only upper limits 
could be measured.
The smallest uncertainty was then tripled and used to calculate 
an abundance based on a 3-$\sigma$ upper limit.

\subsection{Oscillator Strengths}

Table 2 summarizes the oscillator strengths we employed.  Whenever
possible, we have chosen to employ laboratory values.  In the 
appendix, we discuss the effect (usually small) on the abundances that choosing
values from other studies would make, and compare our 
oscillator strengths to those of the Lick-Texas Group (e.g., the linelist
employed in the recent high resolution M3 study by Sneden et al.\ 2004 adopted 
from M5-I01) and
the list of Shetrone \etal\ (2001; dSph-S01) for dSph stars, a list
also incorporated into studies of additional metal-poor dSph stars by Shetrone et al.\ 
(2003) and further discussed by Tolstoy et al.\ (2003) 
and Venn et al.\ (2004).

In deriving the abundances of Sc, V, Mn, Co, Cu, Ba, La, and Eu, we
took into account HFS for all of the lines we
employed.  The La and Eu linelists were taken from Lawler, Bonvallet
\& Sneden (2001) and Lawler \etal\ (2001); Ba from Johnson (2002); and
Cu from Simmerer \etal\ (2003).  In Table 3, we present our HFS
information for the lines we analysed in Sc, V, Mn, and Co,
where the integrated line information and references are listed in
Table 2. In the case of Na and Al, our analyses are consistent with
those in the literature where weak lines are used and HFS is ignored.
Nd has eight isotopes and the odd-Z isotopes have HFS. We do not have
information on the HFS, but using the Nd isotopic splittings from Aoki
\etal\ (2001) did not make a difference, so the effect of saturation
on the lines should be small. For all elements, we use van der Waals
damping constants modified by the Uns\"old approximation.


\subsection{Model Atmospheres}

We used Kurucz model atmospheres (Kurucz 1992, 1993)\footnote{Grids of Kurucz model
atmospheres can be downloaded from 
{\sf http://cfaku5.cfa.harvard.edu/grids.html}.} 
with overshooting and assumed that local thermodynamic equilibrium
(LTE) holds for all species.  In the appendix, we 
discuss the effect on our abundance calculations for different 
sets of model atmospheres.  

Reliable photometry does not exist for all of the stars in our 
sample. We instead derived temperatures by ensuring that the
abundance derived for \ion{Fe}{1} lines did not show
any trends with the atomic parameters of the lines employed. 
As shown by M5-I01 and Kraft \& Ivans (2003; Fe-KI03 hereafter), in their sample of 
metal-poor GGCs RGB stars, 
temperatures derived from the 
color-temperature calibrations of Alonso \etal\ (1999) 
seem to be 
generally in good agreement
with the
excitation temperatures derived by spectroscopic means. 
The photometric 
temperatures we calculate for some of the LMC stars 
are generally cooler than those adopted 
here, and this issue is discussed further in the appendix.  We 
estimate our uncertainty in \teff\ to be 150K.

Our log g values are calculated from the following relationship:

\begin{eqnarray}
\log{g} = \log{\frac{M}{M_{\odot}}} 
	- 0.4(M^{\odot}_{\rm bol} 
	- M_{\rm V} - BC)  \\ + 
\nonumber{4\log{\frac{\rm T_{eff}}{\rm T_{eff}^{\odot}}} + \log{g_{\odot}}},
\end{eqnarray}
where the bolometric correction (BC) was obtained by employing
the formulae of Alonso \etal\ (1999).  We 
adopted ${\rm T_{eff}^{\odot}} = 5770$ K, 
$\log{g_{\odot}} = 4.44$ and 
$M^{\odot}_{\rm bol} = 4.72$.
We also adopt 0.85$M_{\odot}$ for the mass of our stars. Of the 
four different (m-M)$_V$ distance modulus calculations presented by LMC-O98, we used 
the apparent distance modulus they derived 
based on matching the color-magnitude 
diagrams of GGCs to those of the LMC clusters.  The values we derive employing 
this method are in good agreement with those predicted by the 12 Gyr, 
Z = $10^{-4}$ isochrones of Bergbusch \& VandenBerg (1992).
In the case of Hodge 11, homogeneous
photometry for the stars in this cluster did not initially exist, 
so we adopted the isochrone-based log g.  Further 
discussion of these issues are contained in the appendix.

We determined the microturbulent velocity ($\xi$)
spectroscopically by ensuring that there were no trends in the
abundance of \ion{Fe}{1} as a function of EW.  \ion{Fe}{2} was used for the 
input metallicity of the model atmosphere, [m/H]. Our final model 
atmosphere parameters are presented in Table 4.

\begin{deluxetable}{lcccc}
\tablenum{4}
\tablewidth{0pt}
\tablecaption{Model Atmosphere Parameters}
\tablehead{
\colhead{Star} & \colhead {\teff} & \colhead{\logg} & \colhead {$\xi$}
& \colhead{[m/H]}
}
\startdata
 NGC 1898\#1 & 4050  & 0.70  & 2.1  &$-$0.80  \\
 NGC 1898\#2 & 4000  & 0.60  & 2.3  &$-$0.80  \\
 NGC 2005\#1 & 4050 & 0.61 & 2.1 & $-$1.30 \\
 NGC 2005\#2 & 4350  & 1.05  & 1.9  &$-$1.30 \\
 NGC 2005\#3 & 4250 & 0.95 & 2.0 & $-$1.30 \\
 NGC 2019\#1 & 4250 & 0.87 & 2.1 & $-$1.10 \\
 NGC 2019\#2 & 4050 & 0.68 & 2.0 & $-$1.10 \\
 NGC 2019\#3 & 3950 & 0.50 &2.2 & $-$1.10 \\
 Hodge 11\#1 & 4300 & 0.66 & 2.2 & $-$2.00  \\
 Hodge 11\#2 & 4200 & 0.50 & 2.0 & $-$2.00 \\
\enddata
\end{deluxetable}

\subsection{Effect of Stellar Parameters on Derived Abundances}

Our abundances are summarized in Tables 5--14.  
The stellar parameters we adopted result in a disagreement between
the abundances of \ion{Fe}{2} and \ion{Fe}{1}, and \ion{Ti}{2} and \ion{Ti}{1}, 
with the lines of the ionized species producing larger values for 
log$\epsilon$(X) by 0.3--0.5 dex.  Non-LTE effects could account for 
the difference, if the neutral species are ``over-ionized'' due
to the models employed (see Fe-KI03 for an expanded discussion). However, the magnitude of the 
over-ionization effect is not predicted to be as large as this (Gratton 
\etal\ 1999, Korn 2004).  Further discussion regarding these
issues is postponed to the appendix.

\begin{deluxetable}{lrcrccr}
\tablenum{5}
\tablewidth{0pt}
\tablecaption{Abundances for NGC1898\#1}
\tablehead{
\colhead{Species} & \colhead {log $\epsilon$} & \colhead
{$\sigma_{\epsilon}$}
 & \colhead{[X/Fe]\tablenotemark{a}} & \colhead{$\sigma_{[X/Fe]}$} 
& \colhead {$\sigma_{lines}$} & \colhead{N$_{lines}$} 
}
\startdata
\ion{O}{1} &     8.13 &  0.16 &   $-$0.05 & 0.32 &   0.07 &   2 \\
\ion{Na}{1} &    4.58 &  0.18 &   $-$0.49 & 0.22 &   0.10 &   2 \\
\ion{Mg}{1} &    6.44 &  0.25 &    0.12 & 0.21 &   0.29 &   2 \\
\ion{Si}{1} &    6.59 &  0.15 &    0.30 & 0.21 &   0.14 &   4 \\
\ion{Ca}{1} &    4.90 &  0.28 &   $-$0.20 & 0.20 &   0.16 &  15 \\
\ion{Sc}{2} &    2.02 &  0.20 &   $-$0.33 & 0.32 &   0.16 &   2 \\
\ion{Ti}{1} &    3.48 &  0.20 &   $-$0.25 & 0.29 &   0.20 &  33 \\
\ion{Ti}{2} &    4.03 &  0.18 &   $-$0.21 & 0.29 &   0.15 &   7 \\
\ion{V}{1}  &    2.24 &  0.30 &   $-$0.50 & 0.31 &   0.14 &  14 \\
\ion{Cr}{1} &    4.19 &  0.29 &   $-$0.22 & 0.23 &   0.25 &   8 \\
\ion{Mn}{1} &    3.69 &  0.17 &   $-$0.44 & 0.17 &   0.08 &   5 \\
\ion{Fe}{1} &    6.26 &  0.19 &    $-$1.26 & \ldots &   0.20 & 145 \\
\ion{Fe}{2} &    6.77 &  0.36 &    0.51 & 0.35 &   0.21 &   6 \\
\ion{Co}{1} &    3.61 &  0.09 &   $-$0.05 & 0.15 &   0.12 &  10 \\
\ion{Ni}{1} &    4.91 &  0.17 &   $-$0.08 & 0.10 &   0.28 &  20 \\
\ion{Cu}{1} &    2.11 &  0.14 &   $-$0.84 & 0.13 &   0.11 &   2 \\
\ion{Y}{2} &     0.76 &  0.19 &    $-$0.22 & 0.15 &   0.22 &   4 \\
\ion{Zr}{1} &    0.99 &  0.34 &   $-$0.35 & 0.36 &   0.11 &   3 \\
\ion{Ba}{2} &    1.17 &  0.37 &    0.30 & 0.27 &   0.27 &   2 \\
\ion{La}{2} &    0.18 &  0.11 &    0.22 & 0.18 &   0.11 &   2 \\
\ion{Nd}{2} &    0.75 &  0.18 &    0.51 & 0.10 &   0.11 &   6 \\
\ion{Eu}{2} &   $-$0.27 &  0.15 &    0.48 & 0.21 &   0.10 &   1 \\
\enddata 
\tablenotetext{a}{[X/Fe] given for each species except for \ion{Fe}{1}, when
[\ion{Fe}{1}/H] is given}
\end{deluxetable}

\begin{deluxetable}{lrcrccr}
\tablenum{6}
\tablewidth{0pt}
\tablecaption{Abundances for NGC1898\#2}
\tablehead{
\colhead{Species} & \colhead {log $\epsilon$} & \colhead
{$\sigma_{\epsilon}$}
 & \colhead{[X/Fe]\tablenotemark{a}} & \colhead{$\sigma_{[X/Fe]}$} 
& \colhead {$\sigma_{lines}$} & \colhead{N$_{lines}$} 
}
\startdata
\ion{Na}{1} &    5.16 &  0.24 &    0.02 & 0.17 &   0.11 &   2 \\
\ion{Mg}{1} &    6.33 &  0.20 &   $-$0.06 & 0.12 &   0.15 &   2 \\
\ion{Al}{1} &    6.05 &  0.18 &    0.77 & 0.16 &   0.10 &   2 \\
\ion{Si}{1} &    6.72 &  0.16 &    0.36 & 0.24 &   0.15 &   3 \\
\ion{Ca}{1} &    5.00 &  0.32 &   $-$0.17 & 0.19 &   0.16 &  13 \\
\ion{Sc}{2} &    2.18 &  0.16 &   $-$0.05 &0.36 &   0.19 &   6 \\
\ion{Ti}{1} &    3.59 &  0.34 &   $-$0.21 &0.26 &   0.22 &  30 \\
\ion{Ti}{2} &    4.13 &  0.22 &    0.01 & 0.29 &   0.29 &   5 \\
\ion{V}{1} &    2.40 &  0.33 &   $-$0.41 &0.27 &   0.15 &  13 \\
\ion{Cr}{1} &    4.29 &  0.31 &   $-$0.19 & 0.20 &   0.16 &   7 \\
\ion{Mn}{1} &    3.72 &  0.18 &   $-$0.48 & 0.14 &   0.13 &   5 \\
\ion{Fe}{1} &    6.33 &  0.18 &    $-$1.19 & \ldots &   0.24 & 131 \\
\ion{Fe}{2} &    6.65 &  0.32 &    0.32 & 0.37 &   0.28 &   4 \\
\ion{Co}{1} &    3.56 &  0.11 &   $-$0.17 &0.15 &   0.18 &   8 \\
\ion{Ni}{1} &    4.93 &  0.17 &   $-$0.13 &0.10 &   0.20 &  17 \\
\ion{Cu}{1} &    2.21 &  0.17 &   $-$0.81 & 0.14 &   0.14 &   2 \\
\ion{Y}{2} &    0.81 &  0.24 &   $-$0.24 & 0.20  &   0.36 &   4 \\
\ion{Zr}{1} &    1.20 &  0.34 &   $-$0.21 & 0.30 &   0.07 &   3 \\
\ion{Ba}{2} &    1.43 &  0.37 &    0.49 & 0.24 &   0.20 &   2 \\
\ion{La}{1} &    0.23 &  0.12 &    0.20 & 0.19 &   0.13 &   2 \\
\ion{Nd}{2} &    0.89 &  0.25 &    0.58 & 0.16 &   0.32 &   6 \\
\ion{Eu}{2} &   $-$0.24 &  0.14 &    0.44 & 0.23 &   0.10 &   1 \\
\enddata 
\tablenotetext{a}{[X/Fe] given for each species except for \ion{Fe}{1}, when
[\ion{Fe}{1}/H] is given}
\end{deluxetable}

\begin{deluxetable}{lrcrccr}
\tablenum{7}
\tablewidth{0pt}
\tablecaption{Abundances for NGC2005\#1}
\tablehead{
\colhead{Species} & \colhead {log $\epsilon$} & \colhead
{$\sigma_{\epsilon}$}
 & \colhead{[X/Fe]\tablenotemark{a}} & \colhead{$\sigma_{[X/Fe]}$} 
& \colhead {$\sigma_{lines}$} & \colhead{N$_{lines}$} 
}
\startdata
\ion{O}{1}  &    7.68 &  0.16 &    0.10 & 0.30 &   0.10 &   1 \\
\ion{Na}{1} &    4.58 &  0.19 &   $-$0.06 & 0.16&   0.07 &   2 \\
\ion{Mg}{1} &    6.00 &  0.28 &    0.11 & 0.24&   0.34 &   2 \\
\ion{Ca}{1} &    4.61 &  0.28 &   $-$0.06 & 0.18 &   0.15 &  14 \\
\ion{Sc}{2} &    1.76 &  0.17 &    0.01 & 0.28 &   0.18 &   5 \\
\ion{Ti}{1} &    3.04 &  0.33 &   $-$0.26 &0.27 &   0.12 &  16 \\
\ion{Ti}{2} &    3.72 &  0.19 &    0.08 & 0.26&   0.21 &   5 \\
\ion{V}{1} &    1.75 &  0.35 &   $-$0.56 & 0.32&   0.14 &  9 \\
\ion{Cr}{1} &    3.64 &  0.38 &   $-$0.34 & 0.26&   0.06 &   4 \\
\ion{Mn}{1} &    3.06 &  0.23 &   $-$0.64 & 0.19&   0.06 &   3 \\
\ion{Fe}{1} &    5.83 &  0.16 &    $-$1.69 & \ldots &   0.19 & 111 \\
\ion{Fe}{2} &    6.21 &  0.33 &    0.37 & 0.37 &   0.17 &   4 \\
\ion{Co}{1} &    3.08 &  0.12 &   $-$0.15 &0.13 &   0.15 &   5 \\
\ion{Ni}{1} &    4.33 &  0.15 &   $-$0.23 & 0.10&   0.29 &  14 \\
\ion{Cu}{1} &    1.51 &  0.12 &   $-$1.01 & 0.14 &   0.07 &   1 \\
\ion{Y}{2} &    0.36 &  0.14 &   $-$0.19 & 0.14&   0.13 &   3 \\
\ion{Ba}{2} &    0.71 &  0.31 &    0.27 & 0.21 &   0.16 &   2 \\
\ion{La}{2} &   $-$0.18 &  0.17 &    0.29 & 0.20&   0.15 &   1 \\
\ion{Nd}{2} &   $-$0.12 &  0.25 &    0.07 & 0.24 &   0.32 &   2 \\
\ion{Eu}{2} &   $-$0.72 &  0.18 &    0.46 & 0.23 &   0.15 &   1 \\
\enddata
\tablenotetext{a}{[X/Fe] given for each species except for \ion{Fe}{1}, when
[\ion{Fe}{1}/H] is given}
\end{deluxetable}

\begin{deluxetable}{lrcrccr}
\tablenum{8}
\tablewidth{0pt}
\tablecaption{Abundances for NGC2005\#2}
\tablehead{
\colhead{Species} & \colhead {log $\epsilon$} & \colhead
{$\sigma_{\epsilon}$}
 & \colhead{[X/Fe]\tablenotemark{a}} & \colhead{$\sigma_{[X/Fe]}$} 
& \colhead {$\sigma_{lines}$} & \colhead{N$_{lines}$} 
}
\startdata
\ion{Mg}{1} &    6.09 &  0.22 &    0.37 & 0.15 &   0.23 &   3 \\
\ion{Ca}{1} &    4.73 &  0.28 &    0.23 & 0.10 &   0.20 &  12 \\
\ion{Ti}{1} &    3.21 &  0.35 &    0.08 & 0.23 &   0.31 &   6 \\
\ion{Ti}{2} &    4.02 &  0.26 &    0.35 & 0.36 &   0.25 &   3 \\
\ion{Fe}{1} &    5.66 &  0.23 &    $-$1.86 & \nodata &   0.27 &  75 \\
\ion{Fe}{2} &    6.20 &  0.37 &    0.54 & 0.46 &   0.37 &   2 \\
\ion{Ni}{1} &    4.37 &  0.23 &   $-$0.02 & 0.16 &   0.45 &   9 \\
\ion{Sc}{2} &    1.66 &  0.14 &   $-$0.12 & 0.34 &   0.08 &   4 \\
\ion{Mn}{1} &    2.96 &  0.43 &   $-$0.57 & 0.31&   0.29 &   2 \\
\ion{Ba}{2} &    0.77 &  0.31 &    0.50 & 0.26 &   0.19 &   2 \\
\enddata
\tablenotetext{a}{[X/Fe] given for each species except for \ion{Fe}{1}, when
[\ion{Fe}{1}/H] is given}
\end{deluxetable}

\begin{deluxetable}{lrcrccr}
\tablenum{9}
\tablewidth{0pt}
\tablecaption{Abundances for NGC2005\#3}
\tablehead{
\colhead{Species} & \colhead {log $\epsilon$} & \colhead
{$\sigma_{\epsilon}$}
 & \colhead{[X/Fe]\tablenotemark{a}} & \colhead{$\sigma_{[X/Fe]}$} 
& \colhead {$\sigma_{lines}$} & \colhead{N$_{lines}$} 
}
\startdata
\ion{Na}{1} &    4.47 &  0.19 &    0.00 & 0.15 &   0.07 & 1 \\
\ion{Mg}{1} &    5.90 &  0.20 &    0.18 & 0.14 &   0.19 &   3 \\
\ion{Ca}{1} &    4.65 &  0.28 &    0.15 &0.13 &   0.17 &  12 \\
\ion{Sc}{2} &    1.51 &  0.17 &   $-$0.22 & 0.25 &   0.21 &   4 \\
\ion{Ti}{1} &    3.08 &  0.34 &   $-$0.05 & 0.24 &   0.14 &  8 \\
\ion{Ti}{2} &    3.55 &  0.19 &   $-$0.07 & 0.25 &   0.28 &   5 \\
\ion{Cr}{1} &    3.80 &  0.38 &   $-$0.01 & 0.25 &   0.39 &   5 \\
\ion{Mn}{1} &    2.93 &  0.30 &   $-$0.60 & 0.24 &   0.23 &   5 \\
\ion{Fe}{1} &    5.66 &  0.21 &    $-$1.86 & \ldots&   0.27 &  91 \\
\ion{Fe}{2} &    6.15 &  0.28 &    0.49 & 0.37 &   0.21 &   6 \\
\ion{Co}{1} &    2.96 &  0.17 &   $-$0.10 &0.14 &   0.07 &   2 \\
\ion{Ni}{1} &    4.24 &  0.18 &   $-$0.15 & 0.13 &   0.35 &  11 \\
\ion{Cu}{1} &    1.57 &  0.17 &   $-$0.78 &0.15  &   0.11 &   2 \\
\ion{Y}{2} &    0.39 &  0.16 &    0.01 & 0.20&   0.14 &   2 \\
\ion{Ba}{2} &    0.72 &  0.38 &    0.45 & 0.29 &   0.21 &   1 \\
\ion{Nd}{2} &    0.03 &  0.13 &    0.39 & 0.18&   0.15 &   3 \\
\enddata
\tablenotetext{a}{[X/Fe] given for each species except for \ion{Fe}{1}, when
[\ion{Fe}{1}/H] is given}
\end{deluxetable}

\begin{deluxetable}{lrcrccr}
\tablenum{10}
\tablewidth{0pt}
\tablecaption{Abundances for NGC2019\#1}
\tablehead{
\colhead{Species} & \colhead {log $\epsilon$} & \colhead
{$\sigma_{\epsilon}$}
 & \colhead{[X/Fe]\tablenotemark{a}} & \colhead{$\sigma_{[X/Fe]}$} 
& \colhead {$\sigma_{lines}$} & \colhead{N$_{lines}$} 
}
\startdata
\ion{O}{1} &    8.08 &  0.09 &    0.41 & 0.27 &   0.00 &   2 \\
\ion{Na}{1} &    4.68 &  0.22 &   $-$0.36 & 0.20 &   0.21 &   2 \\
\ion{Mg}{1} &    6.57 &  0.21 &    0.28 & 0.09 &   0.12 &   2 \\
\ion{Si}{1} &    7.10 &  0.10 &    0.84 & 0.21 &   0.00 &   1 \\
\ion{Ca}{1} &    5.17 &  0.29 &    0.10 & 0.13 &   0.16 &  15 \\
\ion{Sc}{2} &    1.91 &  0.17 &    0.07 &0.27  &   0.21 &   4 \\
\ion{Ti}{1} &    3.86 &  0.34 &    0.16 & 0.25&   0.18 &  21 \\
\ion{Ti}{2} &    3.88 &  0.24 &    0.15 & 0.30&   0.36 &   4 \\
\ion{V}{1} &    2.61 &  0.36 &   $-$0.10 & 0.28 &   0.14 &  13 \\
\ion{Cr}{1} &    4.44 &  0.31 &    0.06 & 0.17 &   0.20 &   7 \\
\ion{Mn}{1} &    3.71 &  0.28 &   $-$0.39 & 0.17 &   0.13 &   3 \\
\ion{Fe}{1} &    6.23 &  0.19 &    $-$1.29 & \ldots &   0.23 & 129 \\
\ion{Fe}{2} &    6.26 &  0.28 &    0.03 & 0.37 &   0.27 &   4 \\
\ion{Co}{1} &    3.63 &  0.16 &    0.00 & 0.14 &   0.20 &   8 \\
\ion{Ni}{1} &    4.94 &  0.21 &   $-$0.02 & 0.11 &   0.37 &  14 \\
\ion{Cu}{1} &    2.21 &  0.19 &   $-$0.71 & 0.13 &   0.14 &   2 \\
\ion{Y}{2} &    0.99 &  0.21 &    0.04 & 0.20 &   0.21 &   2 \\
\ion{Zr}{1} &    1.57 &  0.39 &    0.26 & 0.31&   0.03 &   3 \\
\ion{Ba}{2} &    1.13 &  0.32 &    0.29 & 0.22 &   0.13 &   2 \\
\ion{Nd}{2} &    0.83 &  0.19 &    0.62 & 0.13 &   0.14 &   5 \\
\ion{Eu}{2} &   $-$0.12 &  0.19 &    0.66 & 0.27 &   0.17 &   1 \\
\enddata
\tablenotetext{a}{[X/Fe] given for each species except for \ion{Fe}{1}, when
[\ion{Fe}{1}/H] is given}
\end{deluxetable}

\begin{deluxetable}{lrcrccr}
\tablenum{11}
\tablewidth{0pt}
\tablecaption{Abundances for NGC2019\#2}
\tablehead{
\colhead{Species} & \colhead {log $\epsilon$} & \colhead
{$\sigma_{\epsilon}$}
 & \colhead{[X/Fe]\tablenotemark{a}} & \colhead{$\sigma_{[X/Fe]}$} 
& \colhead {$\sigma_{lines}$} & \colhead{N$_{lines}$} 
}
\startdata
\ion{O}{1} &    7.93 &  0.18 &    0.03 &  0.35 &   0.13 &   1 \\
\ion{Na}{1} &    4.73 &  0.20 &   $-$0.18 & 0.21 &   0.14 &   2 \\
\ion{Al}{1} &    5.07 &  0.18 &    0.02 & 0.19 &   0.10 &   1 \\
\ion{Mg}{1} &    6.60 &  0.17 &    0.44 & 0.07 &   0.08 &   2 \\
\ion{Ca}{1} &    4.97 &  0.29 &    0.03 & 0.19 &   0.24 &  15 \\
\ion{Sc}{2} &    1.87 &  0.17 &   $-$0.20 & 0.33 &   0.12 &   4 \\
\ion{Ti}{1} &    3.57 &  0.33 &    0.00 & 0.29 &   0.16 &  27 \\
\ion{Ti}{2} &    4.16 &  0.23 &    0.20 & 0.33 &   0.30 &   6 \\
\ion{V}{1} &    2.24 &  0.34 &   $-$0.34 & 0.32 &   0.21 &  14 \\
\ion{Cr}{1} &    4.21 &  0.29 &   $-$0.04 & 0.22 &   0.22 &   8 \\
\ion{Mn}{1} &    3.59 &  0.19 &   $-$0.38 & 0.18  &   0.07 &   3 \\
\ion{Fe}{1} &    6.10 &  0.18 &    $-$1.42 & \ldots &    0.23 & 127 \\
\ion{Fe}{2} &    6.58 &  0.40 &    0.48 & 0.41&   0.35 &   4 \\
\ion{Co}{1} &    3.48 &  0.11 &   $-$0.02 & 0.16 &   0.15 &   5 \\
\ion{Ni}{1} &    4.67 &  0.16 &   $-$0.16 & 0.09 &   0.20 &  18 \\
\ion{Cu}{1} &    2.06 &  0.16 &   $-$0.73 & 0.14 &   0.14 &   2 \\
\ion{Y}{2} &    0.97 &  0.45 &    0.15 & 0.40 &   0.00 &   1 \\
\ion{Zr}{1} &    1.28 &  0.42 &    0.10 & 0.40 &   0.06 &   2 \\
\ion{Ba}{2} &    0.88 &  0.32 &    0.17 & 0.19 &   0.05 &   2 \\
\ion{La}{2} &    0.15 &  0.25 &    0.35 & 0.28 &   0.33 &   2 \\
\ion{Nd}{2} &    0.67 &  0.21 &    0.59 & 0.15 &   0.24 &   4 \\
\ion{Eu}{2} &   $-$0.06 &  0.14 &    0.85 & 0.21 &   0.08 &   1 \\
\enddata
\tablenotetext{a}{[X/Fe] given for each species except for \ion{Fe}{1}, when
[\ion{Fe}{1}/H] is given}
\end{deluxetable}

\begin{deluxetable}{lrcrccr}
\tablenum{12}
\tablewidth{0pt}
\tablecaption{Abundances for NGC2019\#3}
\tablehead{
\colhead{Species} & \colhead {log $\epsilon$} & \colhead
{$\sigma_{\epsilon}$}
 & \colhead{[X/Fe]\tablenotemark{a}} & \colhead{$\sigma_{[X/Fe]}$} 
& \colhead {$\sigma_{lines}$} & \colhead{N$_{lines}$} 
}
\startdata
\ion{O}{1} &     7.81 &  0.11 &   $-$0.01 &0.37  &   0.06 &   1 \\
\ion{Na}{1} &    4.90 &  0.19 &   $-$0.03 & 0.21 &   0.04 &   2 \\
\ion{Mg}{1} &    6.41 &  0.17 &    0.23 & 0.08 &   0.08 &   2 \\
\ion{Si}{1} &    6.53 &  0.14 &    0.38 & 0.20 &   0.07 &   3 \\
\ion{Ca}{1} &    4.94 &  0.28 &   $-$0.02 & 0.21 &   0.20 &  14 \\
\ion{Sc}{2} &    1.77 &  0.18 &   $-$0.22 & 0.38 &   0.18 &   4 \\
\ion{Ti}{1} &    3.59 &  0.32 &    0.00 & 0.31 &   0.21 &  29 \\
\ion{Ti}{2} &    4.23 &  0.28 &    0.35 & 0.40 &   0.51 &   7 \\
\ion{V}{1} &     2.32 &  0.32 &   $-$0.28 & 0.33 &   0.17 &  14 \\
\ion{Cr}{1} &    4.24 &  0.32 &   $-$0.03 & 0.25 &   0.28 &   6 \\
\ion{Mn}{1} &    3.49 &  0.17 &   $-$0.50 & 0.19 &   0.14 &   5 \\
\ion{Fe}{1} &    6.12 &  0.18 &    $-$1.40 & \ldots &   0.23 & 129 \\
\ion{Fe}{2} &    6.41 &  0.42 &    0.29 & 0.41 &   0.37 &   4 \\
\ion{Co}{1} &    3.35 &  0.10 &   $-$0.17 & 0.16 &   0.18 &   8 \\
\ion{Ni}{1} &    4.78 &  0.17 &   $-$0.07 & 0.10 &   0.27 &  17 \\
\ion{Cu}{1} &    1.98 &  0.17 &   $-$0.83 & 0.17 &   0.18 &   2 \\
\ion{Zr}{1} &    1.40 &  0.36 &    0.20 & 0.42 &   0.13 &   2 \\
\ion{Ba}{2} &    1.09 &  0.36 &    0.36 & 0.23 &   0.19 &   2 \\
\ion{La}{2} &    0.27 &  0.17 &    0.45 & 0.16 &   0.22 &   4 \\
\ion{Nd}{2} &    0.97 &  0.44 &    0.87 & 0.35 &   0.35 &   4 \\
\ion{Eu}{2} &   $-$0.03 &  0.13 &    0.86 & 0.19 &   0.07 &   1 \\
\enddata
\tablenotetext{a}{[X/Fe] given for each species except for \ion{Fe}{1}, when
[\ion{Fe}{1}/H] is given}
\end{deluxetable}

\begin{deluxetable}{lrcrccr}
\tablenum{13}
\tablewidth{0pt}
\tablecaption{Abundances for Hodge 11\#1}
\tablehead{
\colhead{Species} & \colhead {log $\epsilon$} & \colhead
{$\sigma_{\epsilon}$}
 & \colhead{[X/Fe]\tablenotemark{a}} & \colhead{$\sigma_{[X/Fe]}$} 
& \colhead {$\sigma_{lines}$} & \colhead{N$_{lines}$} 
}
\startdata
\ion{Mg}{1} &    5.84 &  0.24 &    0.47 & 0.16&   0.19 &   3 \\
\ion{Ca}{1} &    4.41 &  0.26 &    0.26 & 0.12&   0.18 &  12 \\
\ion{Ti}{1} &    2.69 &  0.49 &   $-$0.09 & 0.27&   0.25 &   2 \\
\ion{Ti}{2} &    3.48 &  0.41 &    0.60 & 0.42&   0.38 &   1 \\
\ion{Cr}{1} &    3.43 &  0.47 &   $-$0.03 & 0.21&   0.15 &   3 \\
\ion{Fe}{1} &    5.30 &  0.31 &    $-$2.22 & \nodata&   0.26 &  49 \\
\ion{Fe}{2} &    5.41 &  0.18 &    0.11 & 0.40&   0.20 &   3 \\
\ion{Ni}{1} &    3.87 &  0.29 &   $-$0.17 & 0.17&   0.25 &   4 \\
\ion{Sc}{2} &    1.18 &  0.17 &    0.19 & 0.22&   0.25 &   3 \\
\ion{Mn}{1} &    2.65 &  0.64 &   $-$0.53 & 0.51&   0.46 &   1 \\
\ion{Eu}{2} &   $-$0.35 &  0.10 &    1.35 & 0.29&   0.09 &   1 \\
\ion{Ba}{2} &    0.00 &  0.23 &    0.08 & 0.24&   0.13 &   2 \\
\enddata
\tablenotetext{a}{[X/Fe] given for each species except for \ion{Fe}{1}, when
[\ion{Fe}{1}/H] is given}
\end{deluxetable}

\begin{deluxetable}{lrcrccr}
\tablenum{14}
\tablewidth{0pt}
\tablecaption{Abundances for Hodge 11\#2}
\tablehead{
\colhead{Species} & \colhead {log $\epsilon$} & \colhead
{$\sigma_{\epsilon}$}
 & \colhead{[X/Fe]\tablenotemark{a}} & \colhead{$\sigma_{[X/Fe]}$} 
& \colhead {$\sigma_{lines}$} & \colhead{N$_{lines}$} 
}
\startdata
\ion{Mg}{1} &    5.82 &  0.25 &    0.44 & 0.14 &   0.13 &   4 \\
\ion{Ca}{1} &    4.49 &  0.30 &    0.33 & 0.13&   0.12 &  14 \\
\ion{Sc}{2} &    1.11 &  0.11 &    0.00 & 0.24 &   0.18 &   6 \\
\ion{Ti}{1} &    2.80 &  0.47 &    0.01 & 0.25 &   0.21 &   4 \\
\ion{Ti}{2} &    3.35 &  0.16 &    0.35 & 0.24 &   0.23 &   5 \\
\ion{Cr}{1} &    3.46 &  0.50 &   $-$0.01 & 0.30&   0.43 &   4 \\
\ion{Mn}{1} &    2.39 &  0.47 &   $-$0.80 & 0.26&   0.10 &   1 \\
\ion{Fe}{1} &    5.32 &  0.28 &    $-$2.20 &\ldots &   0.21 &  79 \\
\ion{Fe}{2} &    5.53 &  0.25 &    0.21 & 0.43&   0.40 &   5 \\
\ion{Co}{1} &    2.77 &  0.29 &    0.05 & 0.14&   0.08 &   3 \\
\ion{Ni}{1} &    3.92 &  0.27 &   $-$0.13 & 0.15&   0.34 &  11 \\
\ion{Y}{2} &    0.09 &  0.19 &    0.05 & 0.31 &   0.26 &   3\\
\ion{Ba}{2} &    0.02 &  0.25 &    0.09 & 0.26&   0.08 &   2 \\
\enddata
\tablenotetext{a}{[X/Fe] given for each species except for \ion{Fe}{1}, when
[\ion{Fe}{1}/H] is given}
\end{deluxetable}

In Figure 2, we illustrate the changes in 
log~$\epsilon$(X) due to the estimated uncertainties in each 
of the stellar parameters for NGC~1898\#1, a representative star from our 
sample.  For our particular set of \ion{Fe}{1} and \ion{Fe}{2} lines, most of 
the elemental abundances changes behave in a similar way to the
changes observed in \ion{Fe}{1}.  In subsequent discussions, we reference 
all abundances with respect to \ion{Fe}{1}, except for \ion{O}{1}, \ion{Ti}{2}, and \ion{Sc}{2}, 
which are referenced to \ion{Fe}{2}. Our choice of whether to compare
to \ion{Fe}{1} or \ion{Fe}{2} was based on the relative changes when
atmospheric parameters were changed, as were the sizes of possible
NLTE effects and the ratios that have previously been used in the
literature. Our error analysis is described in detail in the 
appendix. The uncertainties in log($\epsilon$) and in [X/Fe],
including
the effects of the model atmosphere uncertainties, are listed
in Tables 5--14, along with the standard error of the sample for the abundances
 derived from different lines of the same element ($\sigma_{lines}$).

\begin{figure}
\epsscale{.80}
\plotone{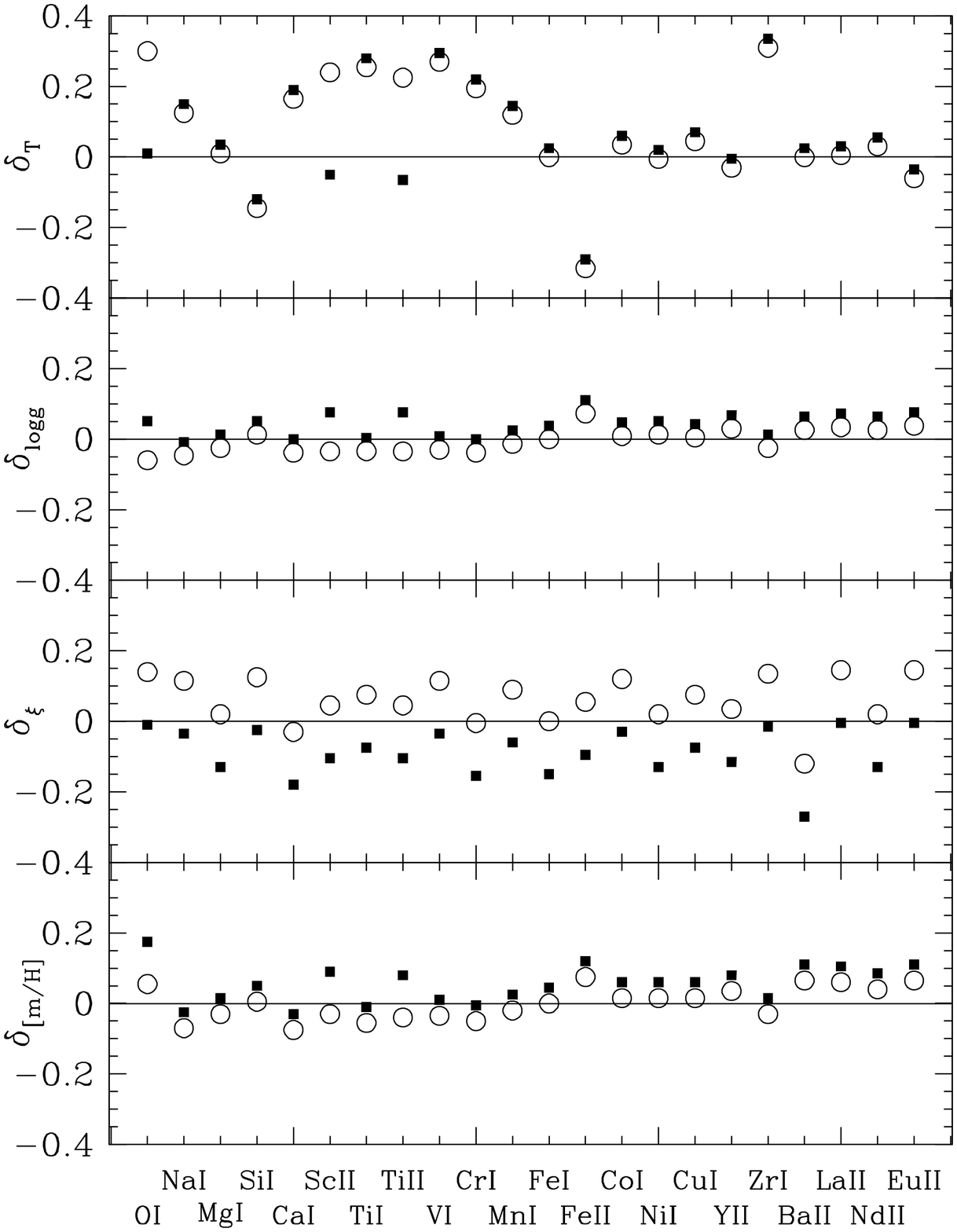}
\caption{The changes in log~$\epsilon$(X) (filled squares)
and [X/Fe] (open circles) for NGC 1898\#1 for
the different species included in our study. The ratio
[X/Fe] has been taken with respect to \ion{Fe}{1} for all
elements except \ion{O}{1}, \ion{Sc}{2}, and \ion{Ti}{2}, which are taken with respect to \ion{Fe}{2}. Fe is taken with respect to H. We varied \teff, 
\logg, $\xi$ and [m/H] individually to produce these plots. The $\delta$ values represpresent changes to \teff{} by 150K, \logg{} by 0.2 dex, $\xi$ by 0.3 
km/s and [m/H] by 0.3 dex. Such dependences as the abundance 
from \ion{Fe}{2} lines on the \teff{} and the abundance from \ion{Ba}{2} 
lines on $\xi$ are readily apparent. The large decrease in the
\ion{Fe}{1} abundance when $\xi$ is increased results in a uniform
increase
in the [X/Fe] values in the third panel.}
\end{figure}

\section{Discussion: Abundance Results}

\subsection{[Fe/H]}

In Table 15, we compare
our values for [Fe/H] in these clusters 
from \ion{Fe}{1} and \ion{Fe}{2} lines with those from the literature.
Our value of [Fe/H] for Hodge 11 is in good agreement with previous literature
estimates, as is the value we derive for NGC 1898. We find both NGC 2005 and NGC 2019 to be more metal-rich than the previous estimates by LMC-O91 from the Ca\thinspace{\sc ii} triplet.  Instead, we derive [Fe/H] values closer to those obtained by LMC-O98 
from the slope of the red giant branch.  
LMC-O91 specifically noted problems with 
their
analysis of these two clusters.  In each case, LMC-O91 could rely on only one star in
the center of the cluster. Our analysis shows that
both of these clusters are closer in metallicity to NGC 1898 than to Hodge 11.
Therefore Figure 17 from LMC-O98 is substantially correct: 
the inner LMC clusters are more metal-rich than the outer LMC clusters
even though both have similar horizontal branch (HB) morphologies. HB morphology is mainly
a function of metallicity, and more
metal-rich clusters have redder HBs. That metallicity
is not the sole variable affecting HB morphology is referred to as
the ``second-parameter problem'' (van den Bergh 1967). In the Milky Way,
Searle \& Zinn (1978) pointed out that the clusters with bluer
HBs (for their metallicity) 
are concentrated within 8 kpc of the Galactic Center, while
the clusters with redder HBs lie at larger distances. With
our confirmation
of the higher metallicities of the inner LMC clusters, 
it appears that the LMC
clusters exhibit second-parameter effects 
that mimic those of the GGCs.

\begin{deluxetable}{lccl}
\tablenum{15}
\tablewidth{0pt}
\tablecaption{[Fe/H] Comparison}
\tablehead{ \colhead {Cluster} & \colhead{[Fe/H]}
 & \colhead{Source} & \colhead{Technique}
}
\startdata
Hodge 11 &  $-$2.21 & This Study &  High-Res Spectra \ion{Fe}{1} \\
         & $-$2.05 & This Study & High-Res Spectre \ion{Fe}{2} \\
         &  $-$2.1$\phn$  & LMC-CH & Low-Res Indices \\
         & $-$2.0$\phn$ & Walker (1993) & Color of RGB \\
NGC 1898 & $-$1.23 & This Study & High-Res Spectra \ion{Fe}{1} \\
         & $-$0.81 & This Study & High-Res Spectra \ion{Fe}{2} \\
         & $-$1.37 & LMC-O91 & Low-Res \ion{Ca}{2} Triplet \\
         & $-$1.18 & LMC-O98 & Slope of RGB \\
NGC 2005 & $-$1.47 & This Study & High-Res Spectra \ion{Fe}{1} \\
         & $-$1.33 & This Study & High-Res Spectra \ion{Fe}{2} \\
         & $-$1.92 & LMC-O91 & Low-Res \ion{Ca}{2} Triplet \\
         & $-$1.35 & LMC-O98 & Slope of RGB \\
NGC 2019 & $-$1.37 & This Study & High-Res Spectra \ion{Fe}{1} \\
         & $-$1.10 & This Study & High-Res Spectra \ion{Fe}{2} \\
         & $-$1.81 & LMC-O91 & Low-Res \ion{Ca}{2} Triplet \\
         & $-$1.23 & LMC-O98 & Slope of RGB \\
\enddata
\end{deluxetable}

\subsection{Literature Sources for Comparison with Abundance Ratios}

One of the primary goals of this paper is to compare the LMC 
cluster abundance ratios with those seen in other stellar populations.
The sample of recent literature studies of the GGCs used for 
comparison in this paper is summarized in Table 16. For Cu, we used
the comprehensive analysis of Simmerer \etal\ (2003) for the GGCs values.
For the elements sensitive to 
proton-capture nucleosynthesis (e.g., O, Na, Mg, and Al), we have 
displayed the abundances from individual stars. For all other elements, as
well
as Mg, we have adopted or calculated the average abundance and 
standard error of the mean (s.e.m.). Using the s.e.m.\
usually gives us a very small ($\sim$0.03) error, much smaller than 
we calculate for the individual errors in the LMC stars. Much of the 
reduction comes from the large number of stars observed in GCC studies.
Some of the reduction, however, is the result of the usually smaller internal 
dispersions in globular cluster abundance ratios where the 
determinations of the relative \teff{}  and \logg\ values are more secure. The [Fe/H] values presented here 
for the GGCs are from Fe-KI03 (Tables 4 and 7) and Kraft \& Ivans 
(2004; Fe-KI04), 
where available, using [Fe/H] derived from Kurucz atmospheres with 
overshooting turned on, corresponding to the choice made for this study. 
For Terzan 7 and NGC 6553, too metal-rich to be included in 
the Fe-KI03 and Fe-KI04 compilations,
we cite the [Fe/H] reported in primary abundance analyses in the literature.
The abundance analyses for 18 of these clusters were done using
evolutionary \logg{} values; the other seven adopted \logg{} based on
ionization equilibrium. In studies where these two methods were
compared (e.g., M5-I01), the \logg{} values for stars at the tip of the 
giant branch from ionization equilibrium constraints are 
$\sim$0.3 dex smaller than those derived from evolutionary considerations. From
Figure 2, this will result in small changes ($< 0.05$ dex) for the majority
of abundance ratios, with the exception of [\ion{Fe}{1}/\ion{Fe}{2}].

\begin{deluxetable}{llllr}
\tablenum{16}
\tablewidth{0pt}
\tablecaption{Literature Sources for Galactic Globular Clusters}
\tablehead{
\colhead{Cluster} & \colhead {Source} & \colhead{[Fe/H]}
& \colhead {\# }}
\startdata
M3  & Sneden \etal\ 2004 & $-$1.42 & 23\\
M4 & Ivans \etal\ 1999 & $-$1.08 &36 \\
M5 & Ivans \etal\ 2001 & $-$1.19 & 36\\
M10 & Kraft \etal\ 1995 & $-$1.43 & 15\\
M13 & Sneden \etal\ 2004 &$-$1.52  & 35\\
M15 & Sneden \etal\ 1997 & $-$2.36  &18 \\
M54 & Brown, Wallerstein,& $-$1.40 & 5\\
 & \& Gonzalez 1999 & & \\
M68 & Lee \etal\ 2004 & $-$2.37 & 7 \\
M71 & Ram\'irez \& Cohen 2002 &$-$0.74 & 25\\
NGC 288 & Shetrone \& Keane 2000 & $-$1.27& 13\\
NGC 362 & Shetrone \& Keane 2000 &$-$1.33 &12 \\
NGC 2808 & Carretta, Bragaglia & $-$1.22& 20\\
 & \& Cacciari 2004 & & \\
NGC 3201 & Gonzalez \& Wallerstein & $-$1.48 &18 \\
 & 1998 &  & \\
NGC 6287 & Lee \& Carney 2002 & $-$2.13& 3 \\
NGC 6293 & Lee \& Carney 2002 & $-$1.97& 2 \\
NGC 6397 & Gratton \etal\ 2001&$-$1.96 & 8 \\
NGC 6541 & Lee \& Carney 2002 &$-$1.76 & 2 \\
NGC 6752 & Gratton \etal\ 2001 &$-$1.49 &18 \\
NGC 6752 & James \etal\ 2004 &$-$1.49 &18 \\
NGC 7006 & Kraft \etal\ 1998 & $-$1.40& 6 \\
NGC 6553 & Cohen \etal\ 1999 & $-$0.18{\tablenotemark{a}}& 5\\
47Tuc & Carretta \etal\ 2004 & $-$0.63 & 12 \\
Pal 5 & Smith, Sneden, \& Kraft & $-$1.34{\tablenotemark{a}} & 4 \\
 & 2002&  &  \\
Pal 12 & Cohen 2004 & $-$0.87 & 3 \\
Terzan 7 & Tautvai{\v s}ien{\.e}\etal\ 2004 &$-$0.61 {\tablenotemark{a}} & 3 \\
\enddata
\tablenotetext{a}{Too metal-rich to have been included by Fe-KI03 or Fe-KI04; adopted from source.}
\end{deluxetable}

We did not apply any corrections to \ion{Na}{1} abundances for non-LTE effects but 
took care to employ a 
similar sample for comparison, revising
the values of two groups by the corresponding cited corrections
(Tautvai{\v s}ien{\.e} \etal\ 2004; Carretta \etal\ 2004b).  On this issue, we refer the reader to discussions by Baum\"uller, Butler, \& Gehren (1998), Gratton \etal\ (1999), Mashonkina, Shimanski\u i, \& Sakhibullin (2000), and Takeda \etal\ (2003).

For field star data, 
we have mainly relied on the work of Fulbright (2000).  
For elements not studied by Fulbright, we have adopted Sc and Co
from Gratton \& Sneden (1991), and Mn from Gratton (1989) and Bai
\etal\ (2004). The data for Cu in field stars is from Mishenina
\etal\
(2002).
All of the additional field star studies for these elements 
include HFS in their analysis. The EWs for V in Fulbright (2000) are
small enough that including HFS effects will not affect the
derived values.
The dSph field star data come from Shetrone \etal\ (2003),
McWilliam, Rich, \& Smecker-Hane (2003), Smecker-Hane \& McWilliam
(2004), Geisler \etal\ (2004), and McWilliam \& Smecker-Hane (2005).

With the exception of iron, our analysis relies on the solar photospheric (where 
reliable) or meteoritic abundances from the compilation of Anders \& 
Grevesse (1989). In the case of iron, we adopt log~$\epsilon$(Fe) = 
7.52, a value close to that recommended by Grevesse \& Sauval 
(1998; log~$\epsilon$(Fe) = 7.50).  We refer the reader to discussions 
by Ryan, Norris \& Beers (1996) and McWilliam (1997), where some of the 
alternative solar iron abundance choices are summarized.  In cases where the 
literature studies employed a solar value different from that adopted 
in this study, we adjusted their abundances to our system.  The 
case where this revision made the most noticeable difference to our 
comparison was to the Co values of Gratton \& Sneden (1991).
When applied to solar EWs, the linelist we employ in this study
reproduces within acceptable errors the solar abundances we adopted 
for all elements.

\subsection{O, Na, Mg, Al}

In all GGCs for which star-to-star abundance variations
have been investigated for correlations between the abundances of
elements sensitive to proton-capture nucleosynthesis, they have been
found.  In Figure 3, we display the abundances of [O/Fe] vs.\ [Na/Fe]
and [Na/Fe] vs.\ [Al/Fe] for our LMC cluster stars along with results
from GGCs.  Very low O/high Na stars do not exist in our sample.  Two
of our stars, NGC 1898\#2 and NGC 2019\#2, have only upper limits for
oxygen.  However, NGC 1898\#2 is also Al-rich, and thus may be a candidate star
for deep mixing.  Overall, the bright giants in our sample of LMC
clusters resemble stars in clusters like M3 and the halo field, rather
than stars in clusters like M13. It is unclear if we should even
expect to see Na values as extreme as those in some of the GGCs -- the
LMC stars may have been born with lower [Na/Fe] values.  Smith \etal\
(2002) found that the LMC stars, in general, had lower [Na/Fe] values
than those in the Milky Way field.

\begin{figure}
\epsscale{.80}
\plotone{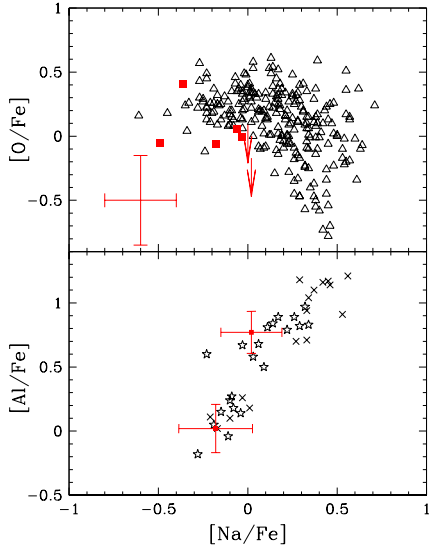}
\caption{ (top) [Na/Fe] vs.\ [O/Fe] for individual stars in 
the LMC (solid squares and limits) 
and the GGCs (open triangles; see Table 16 for sources). 
A typical error bar is shown the lower left.
(bottom) [Al/Fe] vs.\ [Na/Fe] for LMC stars (solid squares) compared
with M3 stars (stars) and M13 stars (crosses). Two LMC
stars from our sample possess both Al and Na measurements.}
\end{figure}

\subsection{$\alpha$ elements}

As seen in Figure 3,  [O/Fe] in most of our sample is at the lower 
boundary of [O/Fe] observed in GGC stars.
In Figure 4, we show the ratios for other $\alpha$ elements, [Mg/Fe], [Si/Fe], 
[Ca/Fe], and [Ti/Fe], compared to stars of the
Galactic halo field, LMC, the dSphs and GGCs.  
While the [Mg/Fe] and [Si/Fe]-ratios we derived are in good 
agreement with those found for field and GGC stars, both the [Ca/Fe] 
and [Ti/Fe] ratios are systematically lower, with a distribution much more 
similar to the abundances found in some of the dSph stars.

\begin{figure}
\includegraphics[width=2.5in,angle=270]{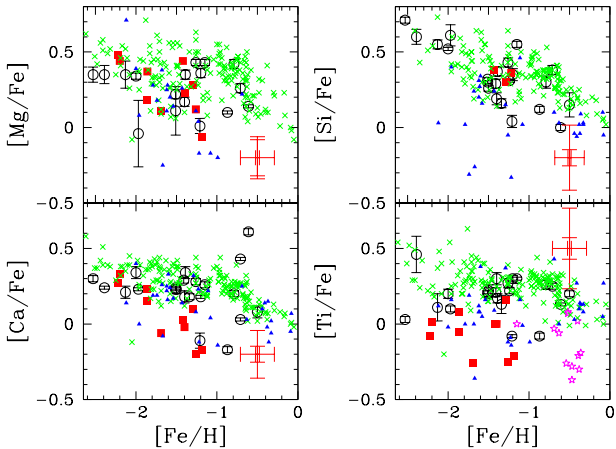}
\caption{[Mg/Fe], [Si/Fe], [Ca/Fe], and 
[Ti/Fe] vs.\ [Fe/H] for individual stars in the LMC clusters 
(solid squares) and dSph (solid triangles), for field stars in 
the Milky Way (crosses), for stars in GGCs (open circles),
and for the LMC field (stars; Smith \etal\ 2002). 
The double error bar represents our estimate of total and random
errors. The larger
errorbar represents the uncertainty including atmosphere parameter
uncertainties, and the smaller the addition of the 
s.e.m. from the lines of the two elements added in quadrature
for the vertical axis or just the s.e.m. for the horizontal. The
latter errorbars are similar to the errorbars of the GGCs. 
In the case of the larger Ti errors, the Ti line strengths, and thus derived abundances, are more 
temperature-sensitive than the other elements.}
\end{figure}

\subsection{Iron-peak elements}

Two general abundance trends are observed in the iron-peak 
elements for Galactic field stars. 
At [Fe/H] $< -2.5$, there are trends with 
[Cr/Fe] and [Mn/Fe] decreasing as with decreasing [Fe/H]. For higher metallicities (and the only metallicities
for which there are GGCs), all the iron-peak ratios, [Sc, V, Cr, Mn, Ni/Fe], 
are essentially solar and show no trend with [Fe/H]. A few
studies have argued for deviations from solar ratios, such
as the supersolar [V/Fe] found in some thick disk stars 
(Prochaska \etal\ 2000), and the low [V/Fe] found in GGC Pal 12 stars
(Cohen 2004).

Abundances of [Co/Fe] and [Ni/Fe], the heavier iron-peak elements, are displayed in Figure 5.
The [Ni/Fe]-values we derive in the LMC are lower than the
average values within the GGCs, and are comparable to the lower Ni
values observed in some dSph and low-Ni GCC stars. However, individual values agree, within
the errors, to those of the GCC cluster and field stars.  
The abundance we derive for Co for the LMC clusters is offset from
most of the GCC abundances displayed in Figure 5. However, as the errorbar
illustrates, the offset could be the result of systematic choices in
the analyses.  We have  excluded the Co abundances from the 
following discussion.  We do, however, note the 
disagreement between the abundances reported for 
metal-poor field and GCC stars as well as the need for additional 
measurements of Co in metal-poor field stars.

\begin{figure}
\includegraphics[width=1.5in,angle=270]{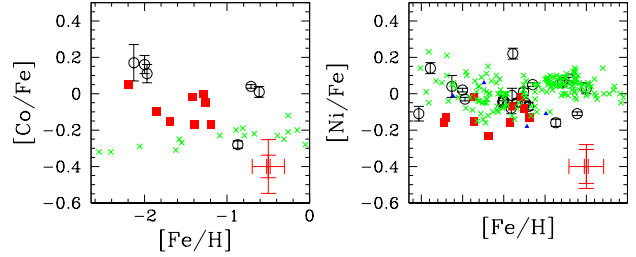}
\caption{[Co/Fe] and [Ni/Fe] for individual stars in the LMC
clusters (solid squares) compared with MW field stars (crosses) and
average values in GGCs (open circles with error bars). A typical error bar for
our data is shown.  As noted in \S~4.5, there is a paucity of comparison 
data for Co in the metallicity 
range of interest here.}
\end{figure}

Figure 6 shows the abundances of the lighter iron-peak elements, [Sc/Fe], [V/Fe], [Cr/Fe], and [Mn/Fe]. While [Sc/Fe], [Cr/Fe] and [Mn/Fe] all
agree with the trends seen in the Milky Way, both field and clusters,
the [V/Fe] ratio in the LMC clusters is decidedly lower. [V/Fe] is
very sensitive to temperature, but so are other ratios in this group,
such as [Cr/Fe]. Even by experimenting with the stellar 
parameters, we cannot find temperatures that force all the
iron-peak ratios to have solar values.  This aspect is illustrated further in the appendix.

\begin{figure}
\includegraphics[width=2.5in,angle=270]{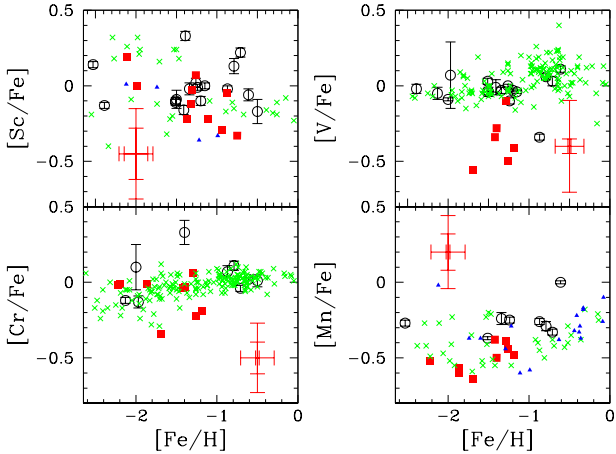}
\caption{Abundance ratios for the iron-peak elements in individual
stars in the LMC clusters (solid squares) compared with MW field
stars (crosses), and average values in GGCs (open circles with error bars). A typical
error bar for our data is shown. With the prominent exception
of V, the iron-peak trends are in good agreement with the Milky
Way trends.  V, like Ti, is very temperature-sensitive but the error bars displayed here incorporate {\it both} random and systematic uncertainties.}
\end{figure}

\subsection{Cu}

The sites for Cu production are thought to be massive stars either through 
the weak $s$-process or Type II SNe, or lower mass binaries through Type Ia SNe
(e.g., Matteucci \etal\ 1993).
Whatever the relative contributions of these
processes, they are also responsible for the strong
trend seen in [Cu/Fe] vs.\ [Fe/H] in the field stars (Sneden \& Crocker
1988; Mishenina \etal\ 2002), a trend also 
shared by the GGCs (Simmerer \etal\ 2003; their Figure 6), but not by the 
individual stars in  \wcen{} (Cunha \etal\ 2002). 
As shown in our Figure 7,
the LMC clusters, like \wcen, show no trend in [Cu/Fe] with
[Fe/H].  The Cu abundances and behavior with [Fe/H] in the stars of \wcen, the LMC, and the dSph systems all appear to be similar to each other but significantly different from the trends and values observed in Milky Way field star or GCC populations.

\begin{figure}[htb]
\epsscale{1.0}
\plotone{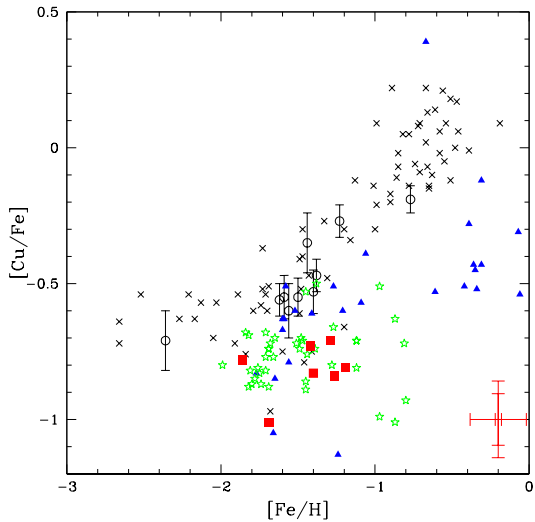}
\caption{[Cu/Fe] vs.\ [Fe/H] for individual stars of the LMC (solid squares) and
dSph systems (solid triangles), in the field of the Milky Way (crosses), the GGCs (open
circles) and
\wcen{} (stars; Cunha \etal\ 2002). The GGC Cu 
values are those plotted in Figure 6 of Simmerer \etal\ (2003). 
Note that the halo field 
stars have slightly larger error bars than those of our study (see
Figure 4).}
\end{figure}

\subsection{Neutron-capture elements}

In the solar system, Eu was
mostly made in the $r$-process, 
while more than 50\% of Y, Zr, Ba, La,
and Nd came from the $s$-process.
These two processes occur, and contribute to the interstellar mix out of which subsequent generations are formed, on
different timescales. While the site of the $r$-process is not
yet definitively known, 
the oldest stars in the Galaxy show the signature of
$r$-process contributions, so a neutron-rich environment associated with 
with massive stars is likely (Truran 1981). The
$s$-process, however, is known to occur in AGB stars, and most efficiently in the
lower-mass AGB stars (see e.g., the review by Busso, Gallino, \& Wasserburg 1999 and references therein). Therefore, $s$-process material should appear
in our clusters if they were formed out of material that had
incorporated the yields of stars which had formed 
at least 1 Gyr previously. 
We measured six neutron-capture elemental abundances in both NGC 2019 and
NGC 1898, and at least one neutron-capture element abundance in the other two clusters. 

Figure 8 compares the abundance ratios
for Y, Ba and Eu in the LMC clusters 
with those from the GGCs, the Milky Way field and the dSphs. First, the LMC
clusters agree in general with the Milky Way field in all three
ratios. Second, the Milky Way system shows large variations in [Ba/Fe], as
M4-I99 and M5-I01 show. Our smaller LMC sample does
not include any examples of the extreme cases observed in the GGCs.
Venn \etal\ (2004) noted that the dSph systems have a large number of stars
with [Y/Fe] substantially lower than the Galaxy and argued that
this was due to a lower contribution of Y (but not Ba or Eu) from the $r$-process. 
Figure 8 shows that the LMC clusters do not share that trend.

\begin{figure}
\epsscale{.80}
\plotone{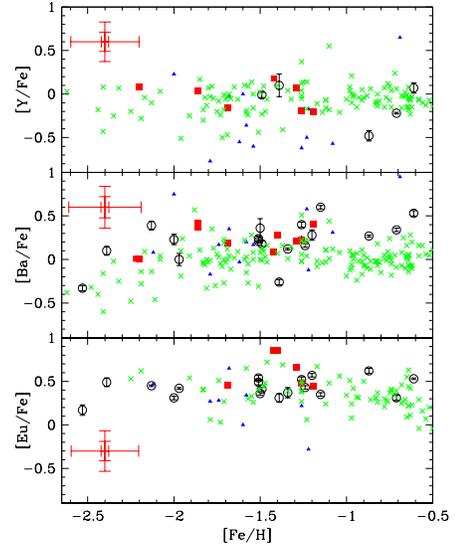}
\caption{The abundance of three neutron-capture elements
in giants in the LMC clusters (solid squares) compared with GGCs (open
circles), and
Galactic field stars (crosses), and dSph stars (triangles). 
The LMC clusters have similar
enhancements to the GGCs and field stars in [Y,Ba,Eu/Fe]. }
\end{figure}

\section{The Chemical History of the LMC Clusters}

The LMC clusters in our sample do not show the same abundance ratios
that we see in the majority of the GGCs. 
To determine if Type Ia SNe ejecta could be an explanation for
the observed LMC [$\alpha$/Fe] patterns, 
we compared the chemical abundances of three stars in two LMC 
clusters to supernova model yields.  Table 17 lists the results, 
along with a subset of those obtained for other stars employing the same techniques, adapted from Table 12 of Ivans et al.\ (2003).  The value of
$<$$N_{Ia}/N_{II}$$>$ represents the 
ratio of the number of SNe Ia to SNe II events that fit both the 
observations and the synthesized mass of Na, Mg, Si, and Fe from 
the model yields.  Two sets of $<$$N_{Ia}/N_{II}$$>$ are shown, one for
each of the two main sources of SNe Ia yields adopted.  The SNe II
yields are taken from Iwamoto \etal\ (1999; see references therein),
integrated over a Salpeter (1955) initial mass function with 
SNe II progenitor star masses from 10--50~M$_{\sun}$.  We refer 
the reader to Ivans et al.\ (2003) for further details regarding
the methods and techniques employed.

\begin{deluxetable}{lccl}
\tablenum{17}
\tablewidth{0pt}
\tablecaption{SNe Ratios Derived from Stellar Abundance Fits to SN Model
Yields}
\tablecolumns{4}
\tablehead{
\colhead{Population} &
\multicolumn{2}{c}{$<$$N_{Ia}/N_{II}$$>$} &
\colhead{Ref.} \\
\colhead{} &
\colhead {Source\tablenotemark{(a)}} &
\colhead {Source\tablenotemark{(b)}} &
\colhead{}
}
\startdata
NGC 1898            & 0.09 $\pm$ 0.07 &  0.08 $\pm$ 0.07 & 1 \\
NGC 2019            & 0.04 $\pm$ 0.03 &  0.04 $\pm$ 0.03 & 1 \\
NGC 6287            & 0.08 $\pm$ 0.18 &  0.08 $\pm$ 0.18 & 2, 3 \\
NGC 6293            & 0.08 $\pm$ 0.18 &  0.08 $\pm$ 0.18 & 2, 3 \\
NGC 6541            & 0.00 $\pm$ 0.03 &  0.00 $\pm$ 0.03 & 2, 3 \\
NS97 Halo Stars     & 0.08 $\pm$ 0.03 &  0.07 $\pm$ 0.03 & 4, 5 \\
NS97 Low-$\alpha$ Stars  & 0.23 $\pm$ 0.11 &  0.20 $\pm$ 0.10 & 4, 5 \\
Sun                 & 0.22 $\pm$ 0.05 &  0.18 $\pm$ 0.01 & 5, 6 \\
Rup 106             & 0.39 $\pm$ 0.32 &  0.34 $\pm$ 0.31 & 5, 7 \\
Pal 12              & 0.38 $\pm$ 0.22 &  0.29 $\pm$ 0.18 & 5, 7 \\
BD+80 245           & 0.58 $\pm$ 0.21 &  0.47 $\pm$ 0.21 & 5 \\
\enddata
\tablenotetext{(a)}{Iwamoto et al.\ (1999).}
\tablenotetext{(b)}{H\"oflich, Khokhlov, \& Wheeler (1995);
H\"oflich \& Khokhlov (1996); H\"oflich, Wheeler, \&
Thielemann (1998); Dom\'inguez, H\"oflich \& Straniero (2001);
H\"oflich et al.\ 2002.}
\tablerefs{
1 -- this study;
2 -- Lee \& Carney (2002);
3 -- Ivans (2005);
4 -- Nissen \& Schuster (1997);
5 -- Ivans et al.\ (2003);
6 -- Anders \& Grevesse (1989) with modified iron abundance
        as discussed in Section 4.1;
7 -- Brown et al.\ (1997);
}
\end{deluxetable}

Listed for comparison are the $<$$N_{Ia}/N_{II}$$>$ values derived for 
Galactic stellar populations of comparable iron abundance, including the sample of 
low-$\alpha$ stars uncovered in the local solar neighbourhood by 
NS97, the more metal-poor low-$\alpha$ star BD+80~245, and the low-$\alpha$ clusters, Rup 106 and Pal 12, studied by Brown et al.\ (1997).  
The value we derive for $<$$N_{Ia}/N_{II}$$>$ in the LMC cluster stars 
is comparable to that of the Galactic halo field stars, and 
disimilar to the $<$$N_{Ia}/N_{II}$$>$ found in the Galactic examples
of low-$\alpha$ stars.  Thus, the 
same mixture of Type Ia and Type II ejecta as found in the
Milky Way halo low-$\alpha$ stars cannot explain the abundance ratios of the LMC stars.
Instead, the $\alpha$-element abundance ratios of the LMC 
stars seem to mimic the splintering of the behavior of the 
$\alpha$-elements seen in the recent studies of inner halo 
clusters by Lee \& Carney (2002) and Lee \etal\ (2004) where
[Si/Ti] is as high as $\sim$0.6 dex for some inner halo GGC stars.

As noted in \S1, numerous studies have reported abundance trends 
with apogalactic distance for stars in our Galaxy.  In particular, Lee \& Carney 
and Lee et al.\ (2004) find abundance
trends in $\alpha$-element ratios that they have interpreted as the result
of mass-dependent yields of Type II SNe. In our study, the most interesting
abundance ratios also come from the inner LMC clusters, in part because
our outer LMC cluster (Hodge 11) is also by far the most metal-poor cluster. 
Combining our results with those of other clusters which lie at varying 
distances from the LMC bar (e.g., Hill \etal\ 2004), future studies will be 
able to compare radial trends between the LMC and Milky Way and should prove 
to be very illuminating. 

Different sources of production of
the various iron-peak elements could explain the variation we observe
in the iron-peak abundance ratios of the LMC cluster stars.  
Timmes \etal\ (1995) used the Woosley \& Weaver calculations (1995) for
Type II SNe and the Thielemann, Nomoto, \& Yokoi (1986) calculations for
Type Ia SNe to look at the chemical enrichment of the Galaxy over time.
They concluded that their model 
underproduced the solar abundance of V and suggested that there is
another source of V, perhaps helium detonations in sub-Chandrasekhar
mass models. 
Nakamura \etal\ (2001) and Umeda \& Nomoto (2002) 
studied the creation of the iron-peak elements
in the extremely energetic class of supernovae known as hypernovae.  They found that, since 
V, along with Cr and Mn, is synthesized in incomplete explosive Si-burning, 
less energetic explosions would produce less V, Cr and Mn.
In the case of the LMC data, we have found subsolar ratios of [V/Fe] and [Mn/Fe] with
solar [Cr/Fe].  Thus, we do not favor the hypernovae explanation. Figure 9
compares our [V/Mn] and [V/Cr] values with predictions of the
hypernova models of Umeda \& Nomoto (2002) and the disagreement is 
clear, regardless of the energy of the explosion or the mass of the
progenitor stars.  It could be the case that fewer V-producing Type Ia SNe
exploded in the LMC than in the Galaxy at this metallicity. 
The previous analysis of the $\alpha$-elements (ignoring Ca and Ti) and iron abundance shows that
the LMC
clusters have similar $<$$N_{Ia}/N_{II}$$>$ values to those found in the
Galactic halo field stars of comparable metallicity. So simply invoking more SNe Type Ia's at lower metallicities cannot explain the other abundance pattern behaviors with respect to Type Ia iron contributions.

\begin{figure}
\epsscale{1.0}
\plotone{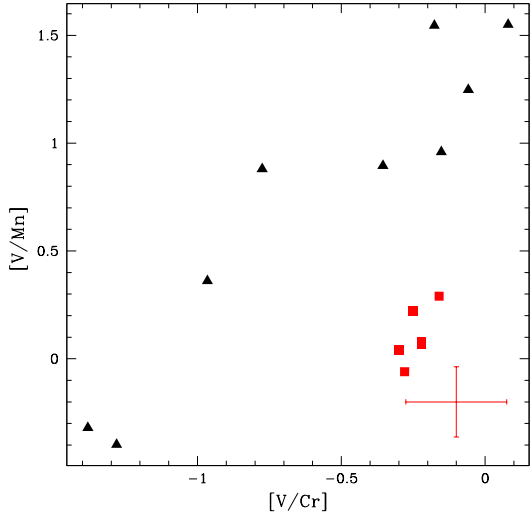}
\caption{Comparison of the [V/Cr] vs [V/Mn] for LMC stars in our
sample (filled squares) with the Umeda \& Nomoto (2002) predictions for the yields of
hypernovae (filled triangles) 
with different progenitor masses (13--30M$_{\odot}$)
and
explosion energies (1--50 $\times 10^{51}$ ergs). These models
were made for zero-metallicity stars, but that should not affect the
results here, since the most important variables in the iron-peak
ratios are the explosion energy and the mass cut. A typical errorbar is
shown for our derived abundances.  The Umeda \& Nomoto hypernovae
yields do not explain the abundances we observe in the LMC cluster
stars.}
\end{figure}

\begin{figure}
\plotone{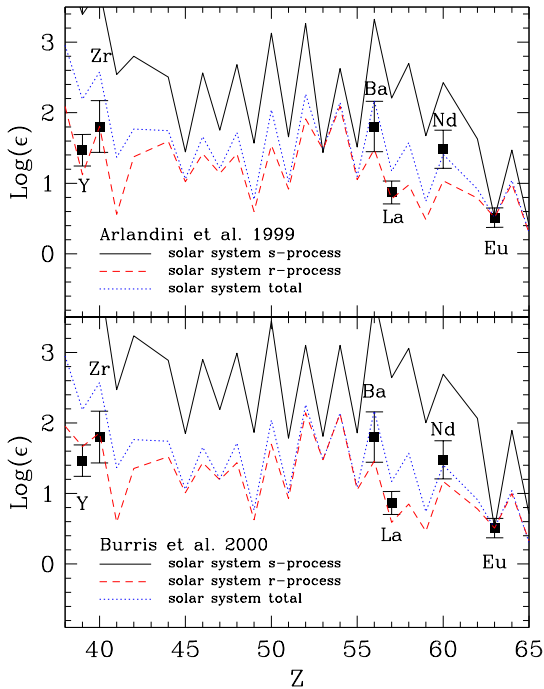}
\caption{Comparison of the average log $\epsilon$ of four LMC
stars with the solar system $r$-process, $s$-process and total fractions
from Arlandini \etal\ (1999) (top) and Burris \etal\ (2000)
(bottom). Because we are comparing results derived for stars with
different metallicities, we first normalized all the log$\epsilon$(Eu)
values to those of NGC 1898\#1. Next,
all the abundances were averaged. The solar values were then adjusted to
match this normalized, averaged Eu value.The agreement between the data and the solar system $r$-process
pattern, especially for Eu and La, shows that there is little, if any,
contribution of the $s$-process to the heavy elements abundances
in the LMC clusters. 
}
\end{figure}

Supporting the argument against low-metallicity SNe Type Ia contributions is the difference in abundance trends between iron-peak elemental ratios.  In Galactic fields stars, the abundance of [Cu/Fe] as a function of [Fe/H] begins to rise $\sim$0.5 dex 
in [Fe/H] before the \afe{} ratio (as a function of [Fe/H]) begins to
decrease (where the decrease in [$\alpha$/Fe] as a function of [Fe/H] is
usually taken to be the effect of Fe contributions from Type Ia SNe 
ejecta).  
While some
Cu can be produced in Type Ia, we favor a metallicity-dependent yield
from Type II's as the best explanation of 
the Galactic data for [Fe/H] $< -1$ as 
predicted by the models of Woosley \& Weaver (1995) and in Timmes \etal\ (1995). The abundances of the LMC cluster and \wcen{} stars (see Figure 7) do not appear
to require much contribution
by Type Ia supernovae. Since the
production of Cu in the LMC matches the production of Fe, we conclude
that there is a minimum [Cu/Fe] produced in another source, possibly the result of massive star contributions.
The lack of any trend in [Cu/Fe] with
metallicity in the LMC and \wcen{} could 
be due to continued contributions from metal-poor SNII or the lack of
contributions from Type Ia SNe. For the LMC clusters, either of these ideas is in accord with the
old ages previously measured. Thus, these results support the idea that the main
$s$-process
is not likely to be a substantial contributor to Cu production (Matteucci \etal\
1993).  However, in the case of the LMC, it may have experienced a history similar to that of the Sgr dSph (see Figure 7).  More relatively metal-rich LMC stars need to be observed in order to discern whether or not a rise in the Cu abundance exists with respect to the Fe abundance.

We explored 
the origin of the neutron-capture elements in our clusters by 
comparing the $r$-process and $s$-process contributions
to the solar system abundances with the observed LMC abundances 
(Figure 10). Our La and Eu abundances are the most reliable, since their
lines are mostly on the linear part of of the curve of growth 
and both HFS and isotopic splitting
are fully taken
into account. 
Viewing the fit to La and Eu (and the other neutron-capture
elements), we favor an $r$-process contribution with, at most, a 20\%
contribution from the $s$-process to improve the agreement with Ba, Nd, and
Y. Thus, the LMC clusters have little, if any, contribution from
the main $s$-process, while the more metal-rich \wcen{}
stars  are $s$-process-rich (e.g., Smith \etal\ 2000), yet both share
the groups of stars low with [Cu/Fe] ratios.

The low $s$-process abundances are a natural consequence of the
LMC clusters being among the first objects formed in the Clouds, before
much chemical evolution took place, and before AGB stars had time to evolve and contribute their yields to the interstellar mix.
We have argued in this section that the addition of iron from
Type Ia ejecta is not the solution to the abundance ratios that we
see, though the lack of Type Ia contributions may be part of the reason for the low [Cu/Fe]
and [V/Fe].

\section{Summary and Conclusions}

Employing a linelist of $>$300 lines with laboratory-based 
gf-values, we have derived elemental abundances for the $\alpha$-, 
iron-peak, and neutron-capture element groups of ten giant stars 
in four old globular clusters of the LMC.  In deriving the 
abundances of Sc, V, Mn, Co, Cu, Ba, La, and Eu, we took into 
account HFS in all of the lines employed.  Extensive numerical 
experiments were performed to elucidate the effects of differing 
choices of gf-values, stellar atmospheres, and stellar parameters 
on the abundances we derived.

While we find that many abundance similarities exist between 
the globular cluster stars in the LMC and our Galaxy (e.g., 
the ratios of [O/Fe], [Na/Fe], [Al/Fe], [Mg/Fe], and [Si/Fe]), 
the same is not true of all of the elements we studied.  In 
particular, we find differences {\em within both} the 
$\alpha$-element {\em and} iron-group abundances.  We find 
lower-than-MWG-average values, and indeed, in some cases, 
clearly sub-solar values of [Ca/Fe], [Ti/Fe], [V/Fe], [Ni/Fe].  
The LMC cluster giant star abundances of [Co/Fe], [Cr/Fe] and 
[Mn/Fe] may indicate additional offsets.  However, the 
available literature on the Galactic abundances of this trio 
of iron-peak elements is sparse (in one or the other of the 
globular cluster or field star populations) and further data
are required to determine with greater certainty whether the
differences in the abundances derived for the different 
groups are significant.

In the case of another iron-peak element, the behavior of 
[Cu/Fe] in our LMC clusters with respect to [Fe/H] appears to 
be constant with a value of $\sim$$-$0.8.  While this is in 
marked contrast to the abundances observed in other MWG halo 
field and cluster stars, it does resemble the trend observed 
in \wcen.  More relatively metal-rich LMC stars need to be 
observed in order to discern whether or not a rise in the Cu 
abundance exists with respect to the Fe abundance

With regards to the neutron-capture elemental abundance 
ratios of [Y/Fe], [Ba/Fe], and [Eu/Fe], we find the LMC 
star results to be similar to  the MWG-average 
values.  We compared the abundances derived for NGC1898 and
NGC2019 against predictions of the scaled solar system 
contributions of the $r$- and $s$-process by Arlandini et al.\
(1999) and Burris et al.\ (2000).  We find that the 
abundances of neutron-capture elements Y, Zr, Ba, La, Nd, and
Eu can largely be accounted for by the $r$-process, with, at
most, a 20\% contribution from the $s$-process to improve the 
agreement with Y, Ba, and Nd.

The abundance ratio distributions observed in red giant stars in the
LMC globular clusters are markedly different from those found in the
GGC red giants, the halo field red giants, and the red giant stars of
the dSph systems.  Since the $\alpha$ elements in the LMC clusters are
not universally suppressed, and the ages of these clusters are old, we
do not favor contributions by Type Ia SNe, but rather a unique star
formation history that produced smaller amounts of Ca, Ti, V, Ni and
Cu than in the Milky Way.  Possible explanations include a bias in
the mass function of SNe that either exploded or whose ejecta were
retained, or stars in the LMC being formed from material resulting from
contributions by lower metallicity SNe than in our Galaxy. The cause
of the low [Y/Fe] values seen in the stars of dSph systems does not
operate in the LMC clusters, and marks another difference, in addition
to the [Mg/Fe] and [Si/Fe] values, between the LMC and the dSph
systems.  There do not appear to be universal trends among the
satellite galaxies of the Galaxy. We conclude that many of the
abundances in the LMC globular clusters we observed are distinct from
those observed in the Milky Way, and these differences are intrinsic
to the stars in those systems.

\acknowledgments

This research has made use of NASA's Astrophysics Data System 
Bibliographic Services; the SIMBAD database, operated at CDS, Strasbourg, 
France; and the NASA/IPAC Extragalactic Database which is operated by the 
Jet Propulsion Laboratory, California Inst.\ of Technology, under contract 
with NASA.  III is pleased to acknowledge research support from NASA 
through Hubble Fellowship grant HST-HF-01151.01-A from the Space Telescope 
Science Inst., operated by AURA, under NASA contract NAS5-26555. Mike Bolte,
Jon Fulbright, Jim Hesser, Bob Kraft, and Andy  McWilliam are warmly
thanked for their helpful and illuminating discussions as well as for 
readily supplying data in electronic form.

\appendix

\section{Error Analysis}

For all elemental abundances derived in this study, the two major sources of 
uncertainty are the EW measurements and the choice of model atmospheres, both of which
overwhelm those from the \loggf.  As discussed in Section 3.1,
the EW uncertainties can be accounted for by the line-to-line scatter.  For a
given elemental abundance, if a smaller than expected value of 
dispersion resulted from the use of a small number of lines (i.e., 
where the standard deviation of the sample was $< 0.05$ dex), we
derived a more accurate value by calculating the expected uncertainty in 
the EW for each line, and adopting the average uncertainty produced. 
The model atmosphere parameters are not independent and are 
correlated in several different ways. We follow the basic method 
outlined by McWilliam \etal\ (1995) and adapted by Johnson (2002):

{\setlength
\arraycolsep{1pt}
\begin{eqnarray}
\sigma^2_{log\epsilon}= \sigma^2_{EW} +
\left({\partial log \epsilon\over\partial T}\right)^2 
\sigma^2_T + 
\left({\partial log\epsilon\over\partial \mlogg}\right)^2 
\sigma^2_{\mlogg} 
+
\left({\partial log\epsilon\over\partial \mfeh}\right)^2 \sigma^2_{\mfeh}  
+ \left({\partial log\epsilon\over\partial \xi}\right)^2 \sigma^2_{\xi} 
{}
\nonumber\\
{}
+  
 2\biggl[\left({\partial log\epsilon\over\partial T}\right)
\left({\partial log\epsilon
\over\partial \mlogg}\right)\sigma_{T\mlogg} 
+ \left({\partial log\epsilon\over\partial \mfeh}\right) 
	\left({\partial log\epsilon\over\partial 
\mlogg}\right) \sigma_{\mlogg \mfeh}
\nonumber 
{}
\nonumber\\
{}
+ {}
\left({\partial log\epsilon\over\partial \mfeh}\right)\left({\partial 
log\epsilon\over T}\right) \sigma_{T \mfeh} \biggr] 
\nonumber
+ {}
{}\left({\partial log\epsilon\over\partial \xi}\right)\left({\partial 
log\epsilon\over \mfeh}\right) \sigma_{\xi \mfeh} \biggr],
\end{eqnarray}}
where $\sigma_{T \mlogg}$, for example, is defined as 
\begin{equation}
\sigma_{T \mlogg}=\frac{1}{N}\sum_{i=1}^N \left(T_i - \overline{T}\right)
\left(\mlogg_{\it i} - \overline{\mlogg}\right).
\end{equation}

The uncertainties in \teff{} and \logg{}  are clearly correlated, 
not only due to the explicit \logg-\teff{} dependence, but also due to 
the dependence of the bolometric correction on the temperature. Equation
1 reveals the other sources of random error in \logg.  We adopted 
uncertainties 
of 0.1 mag in V magnitude, to account for possible systematics 
unaccounted for in the quoted random uncertainties of 0.05 mag. We estimate the 
uncertainty in the apparent distance modulus as 0.2 mag.  Finally, we 
adopt an uncertainty of 0.05 solar masses for stellar mass. Below, 
we discuss the effect of a possible systematic difference
of 0.25 solar masses, as a result of assuming an RGB mass for a star
which has actually undergone mass loss.

Alonso \etal\ (1999) give separate formulae for the bolometric
correction depending on the temperature. Thus, $\sigma_{T \mlogg}$ 
was not determined analytically. Instead, we 
devised the following method to determine the uncertainty in \logg{} and the 
covariance between \teff{} and \logg. We ran 1000 test cases with 
errors added to \teff, V magnitude, (m-M)$_V$, and BC with their 
appropriate $\sigma$s. A \logg{} was then calculated. The set of 
\logg{} and \teff{} allowed us to find $\sigma_{T \mlogg}$. 

We used \ion{Fe}{2} to determine the model metallicity ([m/H]), which is very
sensitive to \logg{} and \teff{}. There is an insufficiently strong
 correlation 
between excitation potential and EW to produce a noticeable 
correlation in the uncertainties of \teff{} and $\xi$. Our uncertainty
in $\xi$ is 0.3 km/s.  While our \ion{Fe}{2} lines are fairly weak, there is a 
correlation between the value for $\xi$ and [m/H]. Since there is a 
large uncertainty in \ion{Fe}{2} due to EW uncertainties, we performed a similar 
calculation to our $\sigma_{\rm T \mlogg}$ calculation to derive 
$\sigma_{\xi \mfeh}$, as well as $\sigma_{\mlogg \mfeh}$ and 
$\sigma_{T \mfeh}$. The uncertainty in the abundance ratios is not 
necessarily equal to adding the uncertainties in [X/H] in quadrature. 
Instead we use equation A19 from McWilliam \etal\ (1995):

\begin{equation}
\sigma(A/B)^2=\sigma(A)^2+\sigma(B)^2-2\sigma_{A,B}.
\end{equation}
 
where the covariance $\sigma_{A,B}$ has been modified from eq. A20
in McWilliam \etal\ (1995) to take into
account the uncertainty in [m/H]:

{\setlength
\arraycolsep{0pt}
\begin{eqnarray}
\sigma_{A,B}=  
\left({\partial log \epsilon_{A}\over\partial T}\right) 
\left({\partial log \epsilon_{B}\over\partial T}\right) 
\sigma^2_{\mlogg} 
 + 
\left({\partial log\epsilon_{A}\over\partial \mlogg}\right)^2 
\left({\partial log\epsilon_{B}\over\partial \mlogg}\right)^2 
\sigma^2_{\mlogg} 
{}
\nonumber\\
+
\left({\partial log\epsilon_{A}\over\partial \mfeh}\right)^2   
\left({\partial log\epsilon_{B}\over\partial \mfeh}\right)^2 \sigma^2_{\mfeh}
+ \left({\partial log\epsilon_{A}\over\partial \xi}\right)^2  
 \left({\partial log\epsilon_{B}\over\partial \xi}\right)^2 \sigma^2_{\xi}   
{}
\nonumber\\
{}
+  
\biggl[\left({\partial log\epsilon_{A}\over\partial T}\right)
\left({\partial log\epsilon_B
\over\partial \mlogg}\right)  + 
\left({\partial log\epsilon_{A}\over\partial \mlogg}\right)
\left({\partial log\epsilon_B
\over\partial T}\right)  \biggr]
\sigma_{T\mlogg}
{}
\nonumber\\
{}
 +
\biggl[\left({\partial log\epsilon_{A}\over\partial \mlogg}\right)
\left({\partial log\epsilon_B
\over\partial \mfeh}\right)  + 
\left({\partial log\epsilon_{A}\over\partial \mlogg}\right)
\left({\partial log\epsilon_B
\over\partial \mfeh}\right)  \biggr]
\sigma_{\mlogg \mfeh} 
{}
\nonumber\\
{}
+  
\biggl[\left({\partial log\epsilon_{A}\over\partial T}\right)
\left({\partial log\epsilon_B
\over\partial \mfeh}\right)  + 
\left({\partial log\epsilon_{A}\over\partial \mfeh}\right)
\left({\partial log\epsilon_B
\over\partial T}\right)  \biggr]
\sigma_{T\mfeh} 
{}
\nonumber\\
{}
+  
\biggl[\left({\partial log\epsilon_{A}\over\partial \xi}\right)
\left({\partial log\epsilon_B
\over\partial \mfeh}\right)  + 
\left({\partial log\epsilon_{A}\over\partial \mfeh}\right)
\left({\partial log\epsilon_B
\over\partial \xi}\right)  \biggr]
\sigma_{T\mlogg} 
\end{eqnarray}}

\section{Discussion of Model Atmosphere Parameters}

Our choice of model atmosphere parameters was a critical part of
our abundance calculations. For most elements, we measure a sufficient
number of lines with reliable \gfvalues{} to make the uncertainty in the 
effective
temperatures the dominant source of uncertainty. In this
appendix, we discuss more fully the changes in abundance ratios that
occur when different atmosphere parameters are adopted.

\subsection{Effective Temperatures}

\teff{} can be derived in a number of ways. We limit our
discussions here to those derived either photometrically, using 
colours (photometric \teff), or spectroscopically, using the abundance derived for
\ion{Fe}{1} lines with different lower excitation potentials (excitation \teff). In
the main body of this paper, we adopted the latter method also
used by Shetrone \etal\ (2003) in their studies of giants in dSphs, and
by the bulk of the studies done by the Lick-Texas group in their analyses
of GGC stars.

\begin{deluxetable}{lccccccc}
\tablenum{18}
\tablewidth{0pt}
\tablecaption{Photometric \& Spectroscopic Temperatures}
\tablehead{
\colhead{Star} & \colhead {V} & \colhead {V-I} & \colhead{V-I$_0$}
& \colhead {m-M$_V$} & \colhead{\teff{} (Alonso)} 
& \colhead{\teff{} (Houdashelt)}
& \colhead{\teff{} (Spec)} 
}
\startdata
NGC 1898\#1 & 16.588 & 1.614 & 1.522 & 18.69 & 3923 & 4101 & 4050 \\
NGC 1898\#2 & 16.464 & 1.654 & 1.562 & 18.69 & 3889 & 4084 & 4000 \\
NGC 2005\#1 & 16.374 & 1.700 & 1.568 & 18.69 & 3883 & 4091 & 4050  \\
NGC 2005\#2 & 16.924 & 1.561 & 1.429 & 18.69 & 4026 & 4161 & 4350  \\
NGC 2005\#3 & 16.834 & 1.646 & 1.514 & 18.69 & 3931 & 4098 & 4250 \\
NGC 2019\#1 & 16.564 & 1.306 & 1.227 & 18.62 & 4306 & 4415 & 4250 \\
NGC 2019\#2 & 16.465 & 1.582 & 1.503 & 18.62 & 3943 & 4124 & 4050  \\
NGC 2019\#3 & 16.248 & 1.720 & 1.641 & 18.62 & 3821 & 4064 & 3950 \\
\enddata
\end{deluxetable}

In their re-evaluation of the abundance scale of globular
clusters using \ion{Fe}{2}, Fe-KI03 employed photometric
\teff{} values because of concerns with the possibility of overionization 
of \ion{Fe}{1}.
For the RGB stars included in their study, they concluded that
\teff{} derived spectroscopically and photometrically were similar, but
because of the degeneracy between \teff{} and \logg{} on
the RGB, spectroscopic temperatures tended to have greater
uncertainty.
We possess photometry for most of the stars in our sample and can compare the \teff{} derived by different means. To correct 
for reddening, and we use the adopted values E(B-V) from 
O98 for NGC 2019, NGC
1898 and NGC 2005 and Walker (1993) for Hodge 11, combined
with the E(B-V)/E(V-I) value from Sarajedini (1994). We used the 
Cousins/Johnson color transformations of Bessell (1979).  Table
18 shows our calculated \teff{} for the Alonso \etal\ (1999) calibration and
the Houdashelt, Sweigart, \& Bell (2000) calibration. We used the
Ram\'irez and Mel\'endez (2005) color-\teff{} calibration on stars
that were blue enough to be included in their system, and found
\teff s within the range covered by Alonso \etal{} and Houdashelt et al.

It is apparent that for the inner LMC clusters NGC 1898, NGC 2005 and NGC 2019,
the \teff{} predictions from the color-temperature calibrations
are a poor match to those we derive spectroscopically. 
A change of about 0.15 mag in the $V-I$ color
due to either uncertainties in reddening or in photometry would
reconcile the spectroscopic and Alonso photometric \teff{} values. These
stars are saturated in the long exposures of LMC-O98, and because of
charge-transfer efficiency 
effects, the length of exposure has an impact on the photometry.
LMC-O98 added a magnitude-dependent 
correction to put the short exposures  on the same system 
as the long exposures. Unfortunately, 
these clusters have not been successfully 
studied from the ground, therefore an independent 
source of
photometry does not exist 
to check the accuracy of the $HST$ results. 

Could the non-standard abundance ratios we observe be eliminated
by any logical choice of \teff? 
To check this possibility, we performed the following test. 
We interpolated a series of Kurucz model atmospheres
with \teff{} from 3800K to 4500K, \logg\ derived from
Equation 1 for the star NGC 1898\#1, and [m/H] derived from
\ion{Fe}{2} lines. In Figure B1, we plot
several important abundance ratios as a function of
the \teff{} of the model atmospheres.  Many abundance ratios are
very dependent to the adopted \teff, and hotter models for NGC
1898\#1 often produced, for some elements, better agreement with ratios derived in the
GGCs. However, other 
than a desire for standard Galactic GC abundance ratios, there is
no reason to adopt these temperatures and several reasons to avoid this
course.

\begin{figure}
\plotone{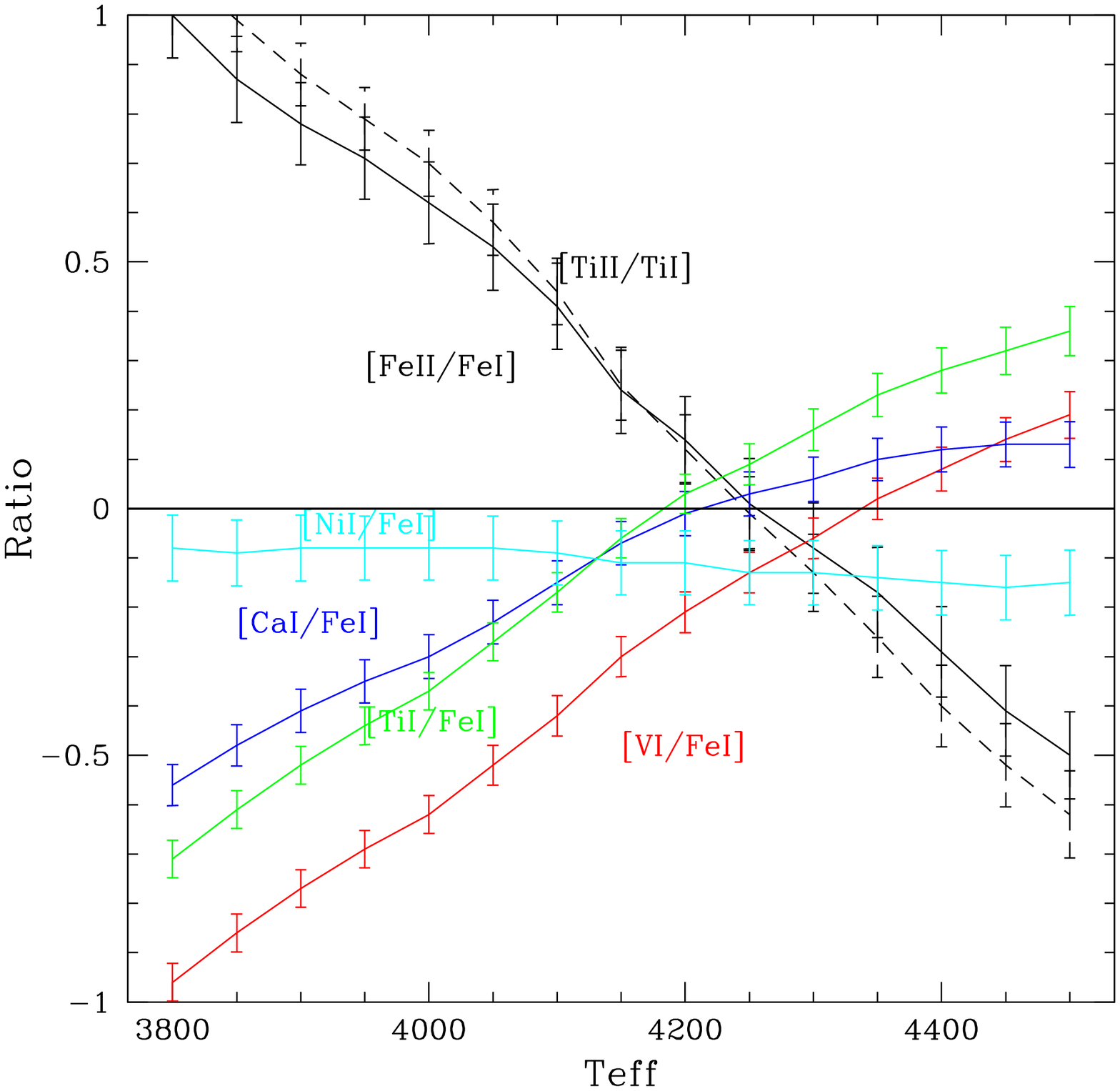}
\caption{Effect on abundance ratios of changing atmospheric parameters
for NGC 1898\#1.
The \teff{} of the model is shown on the horizontal axis, but the
\logg{} and [m/H] are changing as well (see \S~B1). The largest impact on the
abundance ratios are due to changes in \teff{} (see Figure 2). Increasing
the \teff{} improves the agreement between neutral and ionized species
and raises [V/Fe] and [Ti/Fe] closer to solar values. However,
we cannot find a \teff{} that simultaneously makes [Ca/Fe] and [Ti/Fe]
$\sim$0.3 dex and [V/Fe] and [Ni/Fe] $\sim$ 0 (MWG-average values for 
unevolved dwarf stars of comparable metallicity). The temperatures
closest to realizing this are unrealistically too hot for stars on the red giant branch.} 
\end{figure}

First, the spectroscopic and photometric \teff{} values are in 
stark disagreement with the ``chemical temperatures'' derived from
forcing [V/Fe] $\sim$ 0. The reddening of NGC 1898 
would have to be 0.32 mag redder in E(B-V) to make the stars sufficiently
blue to have the photometric \teff{} as high as the chemical \teff{}.
Second, when we compare the chemical \teff{} to the \teff{} derived in
other studies for stars in clusters with similar metallicities, we find
the chemical \teff{} too high for stars  
at the tip of the red giant branch of a
globular cluster. Our derived [Fe/H] and our
adopted
effective temperature are correlated, so raising \teff{}  
decreases the abundance from \ion{Fe}{2} lines. However, the metallicity implied by \ion{Fe}{2}
does not decrease
quickly enough with increasing temperature to reconcile the temperatures.
We conclude that (i) adopting  the Alonso photometric \teff{} would make the
abundance ratios such as [Ca/Fe], [Ti/Fe] and [Sc/Fe], even more
distinct from the standard GGC ratios and (ii) adopting a chemical \teff{} that
force abundance ratios close to those observed in standard GGCs 
would require 
unacceptable reddening and temperatures for the cluster stars.
 
\subsection{Gravities}

In the analysis presented in the main body of the paper, we employed
evolutionary \logg{} values. Many recent studies also use this technique
(see e.g., Cohen et al.\ 2001; M5-I01). It has
the advantage that when choosing the gravity, overionization has no effect
and the adopted metallicity, only a small effect.

A comparison with the Bergbusch and Vandenberg isochrones reveals that
our choice of atmospheric parameters for NGC 1898,
and NGC 2019 cause the stars to occupy a hotter, higher luminosity region than an isochrone with Z = [\ion{Fe}{2}/H] (Figure 2B).
A decrease in the \teff{} we chose could
reconcile the evolutionary and isochronic \logg{} values.
If we changed \logg{} by changing the distance or the bolometric
correction, we need to make a larger correction than shown in Figure
A2.  Because the derived \ion{Fe}{2} is very sensitive to the
adopted \logg, an increase in \logg{} would force the adoption of a
higher
Z isochrone. For the parameters used here, this process converges with
an increase in \logg{} of 0.3 dex and an increase in
[\ion{Fe}{2}/H] of 0.15 dex. 

We also investigated the consequence of assuming
too large a mass for these tip stars. 
If instead of having a mass of 
0.85 M${_\odot}$, these stars may have undergone sufficient mass loss to 
have a typical mass of an HB star (0.6M${_\odot}$). 
Equation 1 shows that the calculated  \logg{} would 
decrease by 0.15 dex. However, for most
cases, the change in the derived abundances is $<$0.1 dex. As shown in
Figure 2, these changes are smaller than those due to temperature
uncertainties.

\begin{figure}
\plotone{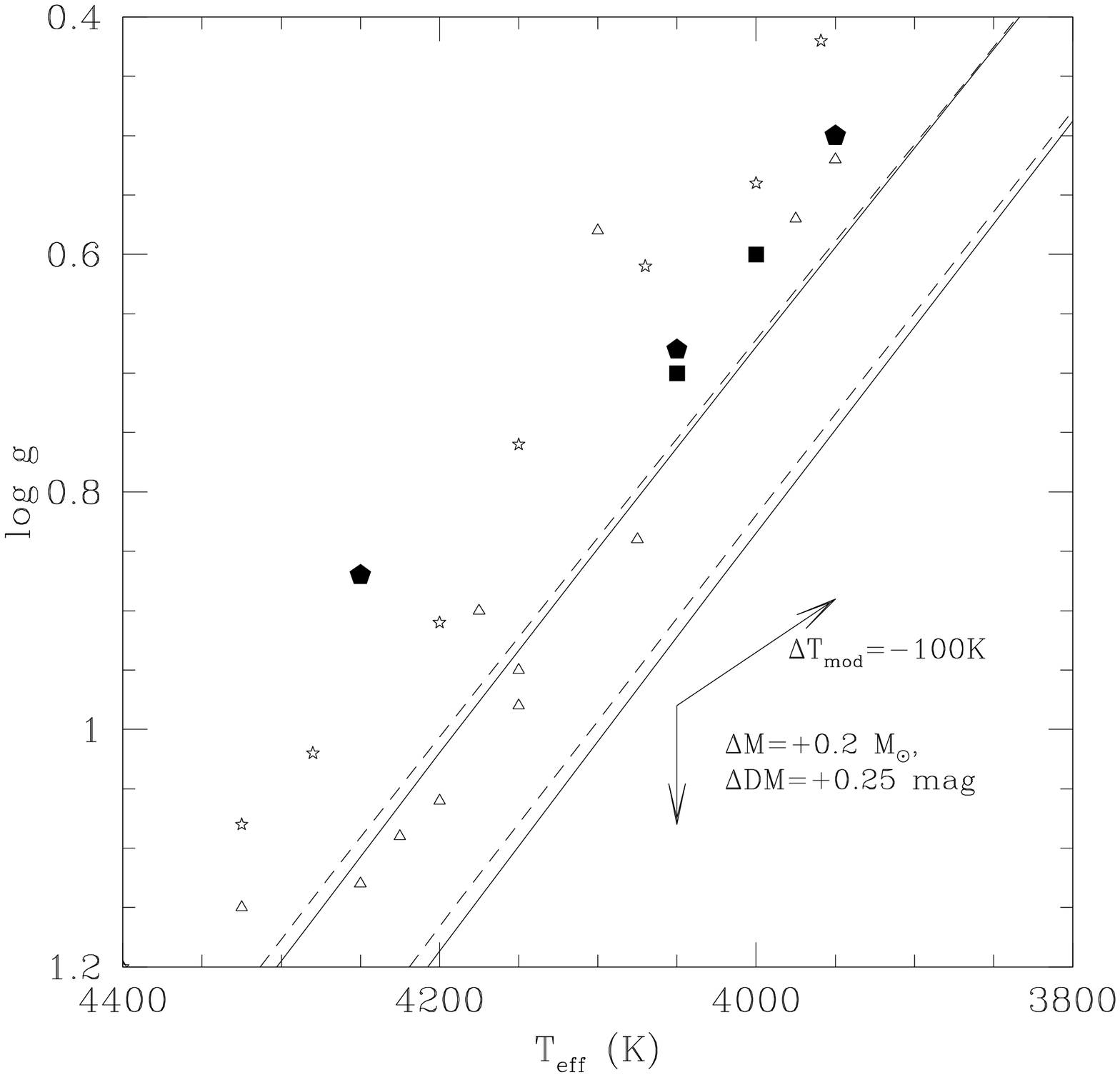}
\caption{Comparison of our adopted \logg{} values with the Bergbusch
\& VandenBerg (1992) isochrones for stars 
in two clusters with similar
metallicities: NGC 1898 and NGC 2019 (filled symbols). The Z = 0.0017 
and 0.001 ([m/H] = $-$1.03 and $-$1.26, respectively) isochrones of
8 Gyr (dashed) and 12 Gyr (solid) are plotted. The more metal-rich
isochrones
lie to the right. Stars from M4-I99
(open triangles) ([Fe/H] = $-$1.15)  
and M5-I01 (open stars) ([Fe/H] = $-$1.24) are also shown. The metallicities
derived from \ion{Fe}{2} lines for the LMC clusters make them 
slightly more metal-rich than M4 or
M5, so their points should lie either to the right or down from the
other clusters. The arrows indicate the magnitude
of the shift in the Kiel diagram if
the temperature of the model (which in turn affects the
log g as shown in Eq. 1), the mass of the star, or the distance modulus
were changed by the indicated amount. } 
\end{figure}

The discrepancy between the abundance from \ion{Fe}{1}
and \ion{Fe}{2} lines could be reduced by adopting a lower
\logg. However, even by reducing the \logg{} to 0, the limit of the
Kurucz grid, we were still $\sim$ 0.1 dex from ionization equilibrium for 
several stars. Eq. 1 shows that decreasing \logg{} by 0.7-1.0 dex would 
result in an unacceptably large increase in the distance modulus to
the LMC of 1.75-2.5 mag. Thus, the evolutionary \logg{} values, along with
the previously published distance estimates, are the most correct 
\logg{} values to employ here.

\subsection{Choice of grid of model atmospheres}

We interpolated model atmospheres using the
 standard grid of Kurucz models with overshooting (Kurucz 1992, 1993).
Other possible
choices include Kurucz models without overshooting (Castelli, Gratton,
\& Kurucz 1997), Kurucz models
without overshooting, but with new opacity distribution function
sampling and enhancements in the $\alpha$-elements
 (Castelli \& Kurucz 2003), an early version of MARCS (Bell
\etal{} 1976)
models, and the updated MARCS spherically geometric models (Gustafsson \etal{}
 2002)\footnote{Grids of the new MARCS  model
atmospheres can be downloaded from 
{\sf http://www.marcs.astro.uu.se}.} . 
Table 19 summarizes the effect on the abundance ratios of
star NGC 2019\#2 for those model atmospheres in the above order.
The second column displays the results found in our study and
the remaining columns display the difference in the abundance 
derived employing an alternate model in the sense of $\Delta$
$\equiv$ (alternative model results) $-$ (this study).

We find a similar result to M4-I99 when comparing the effect of 
the different models, namely that the original MARCS models gave
similar answers to Kurucz models that were 50K hotter. In general,
the choice of the
grid of model atmospheres is a much smaller source of uncertainty than
the choice of \teff.  However, we note the significant revision one
would find if the spherically geometric models were employed in place
of plane-parallel models used in this study (and in all of the 
abundance studies with which we have compared our results against):
the difference between the derived abundances of \ion{Fe}{1} and
\ion{Fe}{2} increases from 0.5 to 0.7~dex.

\begin{deluxetable}{lrrrrr}
\tablenum{19}
\tablewidth{0pt}
\tablecaption{NGC 2019\#2 $\Delta$ Abundances Due To Model Atmosphere Grid Choices}
\tablehead{
\colhead{Species} & \colhead {Kurucz} & \colhead {$\Delta$
  Kurucz}
 & \colhead {$\Delta$Kurucz} & 
\colhead{$\Delta$MARCS} & \colhead{$\Delta$MARCS}
\\
\colhead{} & \colhead {over} & \colhead {nover} & \colhead{$\alpha$}  
&  \colhead{old} 
& \colhead{new} 
}
\startdata
\ion{O}{1}  & 7.93  &    0.00 &   0.08 & $-$0.03 &    0.13 \\
\ion{Na}{1} & 4.73  & $-$0.01 &$-$0.01 &    0.02 & $-$0.09 \\
\ion{Mg}{1} & 6.60  &    0.00 &   0.05 & $-$0.02 & $-$0.01 \\
\ion{Ca}{1} & 4.97  & $-$0.01 &   0.06 & $-$0.06 & $-$0.08 \\
\ion{Sc}{2} & 1.87  &    0.02 &   0.11 & $-$0.05 &    0.16 \\
\ion{Ti}{1} & 3.57  & $-$0.01 &   0.05 &    0.08 & $-$0.08 \\
\ion{Ti}{2} & 4.16  &    0.00 &   0.10 & $-$0.05 &    0.16 \\
\ion{V}{1}  & 2.24  & $-$0.01 &   0.06 &    0.09 & $-$0.06 \\
\ion{Cr}{1} & 4.21  & $-$0.01 &   0.04 &    0.04 & $-$0.07 \\
\ion{Mn}{1} & 3.59  & $-$0.01 &   0.04 &    0.02 &    0.00 \\
\ion{Fe}{1} & 6.10  &    0.00 &   0.07 & $-$0.02 &    0.07 \\
\ion{Fe}{2} & 6.58  &    0.01 &   0.15 & $-$0.10 &    0.28 \\
\ion{Co}{1} & 3.48  &    0.00 &   0.06 & $-$0.02 &    0.07 \\
\ion{Ni}{1} & 4.67  &    0.01 &   0.09 & $-$0.02 &    0.11 \\ 
\ion{Cu}{1} & 2.06  &    0.00 &   0.07 & $-$0.02 &    0.08 \\
\ion{Y}{2}  & 0.73  &    0.00 &   0.09 & $-$0.04 &    0.13 \\
\ion{Zr}{1} & 1.32  & $-$0.02 &   0.09 &    0.13 & $-$0.05 \\
\ion{Ba}{2} & 0.88  &    0.01 &   0.15 &    0.02 &    0.16 \\
\ion{La}{2} & 0.15  &    0.00 &   0.10 & $-$0.04 &    0.16 \\
\ion{Nd}{2} & 0.59  &    0.00 &   0.10 & $-$0.03 &    0.14 \\
\ion{Eu}{2} & $-$0.06 &  0.00 &   0.10 & $-$0.06 &    0.18 \\
\enddata
\tablenotetext{a}{$\Delta$ $\equiv$ (alternative model results) $-$ (this study).}
\end{deluxetable}

\subsection{Oscillator Strengths}

In order to investigate the difference made by using our
linelist vs.\ those of other selected studies, we
performed the following experiment. 
We redid the analysis for star NGC 2019\#2 using the line
lists of M5-I01 and Shetrone \etal\ (2001) and our equivalent widths
and HFS constants. As seen in Table 3B, for many elements the
results do not change by more than 0.05 dex. In the cases with
larger deviations, this is mainly due to the small number of lines and
the uncertainties in the EWs, rather than large changes in the \gfvalues.

\begin{deluxetable}{lrrrrrr}
\tablenum{20}
\tablewidth{0pt}
\tablecaption{Effect of Linelists on Abundances of NGC 2019\#2}
\tablehead{
\colhead{Species} & \colhead {This Study} & \colhead{\# of lines} & 
\colhead {M5-I01} & \colhead {\# of lines} &
\colhead{dSph-S01} & \colhead {\# of lines} 
}
\startdata
\ion{O}{1} & 7.93 & 1  & 7.93 & 1 & \nodata & \nodata \\
\ion{Na}{1} & 4.73 & 2 & 4.73 & 2 & \nodata & \nodata \\
\ion{Mg}{1} & 6.60 &2 & 6.60 & 2 & \nodata & \nodata \\
\ion{Al}{1} & 5.07 & 1 & 5.29 & 1 & \nodata & \nodata \\
\ion{Ca}{1} & 4.97 & 15 & 5.03 & 3 & 5.07 & 3 \\
\ion{Ti}{1} & 3.57 & 27 & 3.61 & 5 & \nodata & \nodata \\
\ion{Ti}{2} & 4.16 & 6 & \nodata & \nodata & 4.33 & 1 \\
\ion{V}{1} & 2.24 & 14 & 2.12 & 1 & 2.27 & 4 \\
\ion{Fe}{1} & 6.10 & 127 & 6.11 & 21 & 6.11 & 27 \\
\ion{Fe}{2} & 6.58 & 4 & 6.39 & 3 & 6.49 & 3 \\
\ion{Ni}{1} & 4.67 & 18 & 4.74 & 2 & \nodata & \nodata \\
\ion{Y}{2} & 0.73 & 5 & \nodata & \nodata & 0.97 & 1\\
\ion{Nd}{2} & 0.67 & 4 &  \nodata & \nodata & 0.54 & 1 \\
\ion{Eu}{2} & $-$0.06 & 1 & $-$0.14 & 1 & $-$0.14 & 1 \\
\ion{Sc}{2} & 1.87 & 4 & 1.97 & 2 & \nodata & \nodata \\
\ion{Mn}{1} & 3.59 & 3 & \nodata & \nodata & 3.59 & 3 \\
\ion{Co}{1} & 3.48 & 5 & \nodata & \nodata & 3.47 & 1 \\
\ion{La}{2} &  0.15 & 2 & 0.38 & 1 & \nodata & \nodata \\
\ion{Ba}{2} & 0.88 & 2 & 0.88 & 2 & \nodata & \nodata \\
\enddata
\end{deluxetable}

\subsection{Summary}

A variety of methods can be used to determine the model atmosphere
parameters of cool giant stars. In the ideal situation, all of the
methods would provide consistent answers. However, this is seldom
the case. In this appendix, we have investigated the effect of
choices in temperature (photometric vs.\ spectroscopic), reddening
and distance modulus, \logg{} and evolutionary state, model 
atmosphere code, and adopted atomic parameters. In no instance do
any of these experiments force us to significantly revise the
abundance ratios derived in this study. Instead, we conclude that
many of the abundances in the LMC globular cluster stars are distinct from those observed
in GGC stars of similar metallicities and these 
differences are intrinsic to the stars in those systems, not
induced by the method of analysis.



\begin{references}

Alonso, A., Arribas, S., \& Mart\'inez-Roger, C.\ 1999, A\&AS, 140, 261

Anders, E.\ \& Grevesse, N.\ 1989, Geochim.\ Cosmochim.\ Acta, 53, 197

Aoki, W., Ryan, S.\ G., Norris, J.\ E., Beers, T.\ C., Ando, H., Iwamoto, 
	N., Kajino, T., Mathews, G.\ J., \& Fujimoto, M.\ Y.\ 2001, ApJ, 
	561, 346

Arlandini, C., K\"appeler, F., Wisshak, K., Gallino, R., Lugaro, M., 
	Busso, M., \& Straniero, O.\ 1999, \apj, 525, 886

Arnett, D. 1971, ApJ, 166, 153

Bai, G.\ S., Zhao, G., Chen, Y.\ Q., Shi, J.\ R., Klochkova, V.\ G.,
	Panchuk, V.\ E., Qiu, H.\ M., \& Zhang, H.\ W.\ 2004, A\&A, 425, 671

Bard, A.\ \& Kock, M.\ 1994, A\&A, 282, 1014

Bard, A., Kock, A., \& Kock, M.\ 1991, A\&A, 248, 315

Baum\"uller, D., Butler, K., \& Gehren, T.\ 1998, \aap, 338, 637

Bell, R.\ A., Eriksson, K., Gustafsson, B., \& Nordlund, A.\ 1976, A\&AS,
	23, 37

Bellazzini, M., Ferraro, F.\ R., \& Ibata R.\ 2003, AJ, 125, 188

Bessell, M.\ S.\ 1979, PASP, 91, 589

Bergbusch, P.\ A., \& VandenBerg, D.\ A.\ 1992, ApJS, 81, 163

Bernstein, R., Shectman, S.\ A., Gunnels, S.\ M., Mochnacki, S., \&
	Athey, A.\ E.\ 2003, Proceedings of the SPIE, 4841, 1694 

Blackwell, D.\ E., Booth, A.\ J., Haddock, D.\ J., Petford, A.\ D., \& Leggett,
	S.\ K.\ 1986, MNRAS, 220, 549

Blackwell, D.\ E., Booth, A.\ J., Menon, S.\ L.\ R., \& Petford, A.\ D.\ 
	1986, MNRAS, 220, 289

Blackwell, D.\ E., Menon, S.\ L.\ R., \& Petford, A.\ D.\ 1983, MNRAS, 
	204, 883

Blackwell, D.\ E., Menon, S.\ L.\ R., \& Petford, A.\ D.\ 1984, MNRAS, 
	207, 533

Blackwell, D.\ E., Menon, S.\ L.\ R., Petford, A.\ D., \& Shallis, M.\ J.\ 
	 1982b, MNRAS, 204, 883

Blackwell, D.\ E., Petford, A.\ D., Shallis, M.\ J., \& Leggett, S.\ K.\ 
	1982a, MNRAS, 199, 21

Blackwell, D.\ E., Petford, A.\ D., Shallis, M.\ J., \& Simmons, G.\ J.\ 
	1980, MNRAS, 191, 445

Blackwell, D.\ E., Petford, A.\ D., Shallis, M.\ J., \& Simmons, G.\ J.\ 
	1982c, MNRAS, 199, 43

Blackwell, D.\ E., Petford, A.\ D., \& Simmons, G.\ J.\ 1982, MNRAS, 201, 595

Blackwell, D.\ E., Ibbetson, P.\ A., Petford, A> D, \& Shallis, M.\ J.\ 
	1979, MNRAS, 186, 633

Blackwell, D.\ E., Petford, A.\ D., \& Shallis, M.\ J.\ 1979, MNRAS, 186, 657

Booth, A.\ J., Shallis, M.\ J., \& Wells, M.\ 1983, MNRAS, 205, 191


Brocato, E., Castellani, V., Ferraro, F.\ R., Piersimoni, A.\ M., \& Testa, 
	V.\ 1996, MNRAS, 282, 614

Brown, J.\ A., \& Wallerstein, G.\ 1992, AJ, 104, 1818

Brown, J.\ A., Wallerstein, G.\ \& Gonzalez, G.\ 1999, AJ, 118, 1245

Brown, J.\ A., Wallerstein, G.\ \& Zucker, D.\ 1997, AJ, 114, 180

Buonanno, R., Buscema, G., Fusi Pecci, F., Richer, H.\ B., \& Fahlman, 
	G.\ G.\ 1990, AJ, 100, 1811

Buonanno, R., Corsi, C.\ E., Fusi Pecci, F., Fahlman, G., \& Richer, H.\ B.\ 
	1994, ApJ, 430, L121

Burris, D.L., Pilachowski, C.A., Armandroff, T.E., Sneden, C., Cowan, 
	J.\ J.\ \& Roe, H.\ 2000, ApJ, 544, 302

Busso, M., Gallino, R., \& Wasserburg, G., J.\ 1999, \araa, 37, 239

Cardon, B.\ L., Smith, P.\ L., Scalo, J.\ M., Testerman, L., \& Whaling,
	W.\ 1982, ApJ, 260, 395

Carretta, E., Bragaglia, A., \& Cacciari, C.\ 2004, ApJ, 610, L25

Carretta, E., Gratton, R.\ G., Bragaglia, A., Bonifacio, P., \&
	Pasquini, L.\ 2004, A\&A, 416, 925 

Castelli, F., \& Kurucz, R.\ L.\ 2003, in Modelling of Stellar Atmospheres, 
	(ed) N.\ E.\ Piskunov, W.\ W.\ Weiss, D.\ F.\ Gray, ASP, 210, 20

Childs, W.\ J., Poulsen, O., Goodman, L.\ S., \& Crosswhite, H.\ 1979,
	Phy.\ Rev.\ A, 19, 168

Cochrane, E.\ C.\ A., Benton, D.\ M., Forest, D.\ H., \& Griffith,
	J.\ A.\ R.\ 1998, J.\ Phys.\ B, 31, 2203

Cohen, J.\ G.\ 2004, AJ, 127, 1545

Cohen, J.\ G., Gratton, R.\ G., Behr, B.\ B., \& Carretta, E.\ 1999, 
	ApJ, 523, 739

Cottrell, P.\ L., \& Da Costa, G.\ S.\ 1981, ApJ, 245, 79

Cowley, A.\ P.\ \& Hartwick, F.\ D.\ A.\ 1982, ApJ, 259, 89

Cunha, K., Smith, V.\ V., Suntzeff, N.\ B., Norris, J.\ E., Da Costa, G.\ S.,
	\& Plez, B.\ 2002, AJ, 124, 379

Den Hartog, E.\ A., Lawler, J.\ E., Sneden, C., Cowan, J.\ J.\ 2003, ApJS, 
	148, 543

Denissenkov, P.\ A.\ \& Herwig, F.\ 2003, ApJ, 590, 99

Denissenkov, P.\ A.\ \& VandenBerg, D.\ A.\ 2003, ApJ, 593, 509

Denissenkov, P.\ A.\ \& Weiss, A.\ 1996, A\&A, 308, 773

Dinescu, D.\ I., Girard, T.\ M., \& van Altena, W.\ F.\ 1999, AJ, 117, 1792

Dinescu, D.\ I., Majewski, S.\ R., Girard, T.\ M., \& Cudworth, K.\ M.\ 
	2000, AJ, 120, 1892

Dom\'inguez, I., H\"oflich, P.\ \& Straniero, O.\ 2001, \apj, 557, 279

Fitzpatrick, M.\ J.\ \& Sneden, C.\ 1987, \baas, 19, 1129

Fuhr, J.\ R., Martin, G.\ A., \& Wiese, W.\ L.\ 1988,
	J.\ Phys.\ Chem.\ Ref.\ Data, 17, 4

Fulbright, J.\ P.\ 2000, AJ, 120, 1841

Fulbright, J.\ P.\ 2002, AJ, 123, 404

Fulbright, J.\ P.\ 2004, in Carnegie Observatories Astrophysics Series, 
	Vol.\ 4: Origin and Evolution of the Elements, ed.\ A.\ McWilliam 
	and M.\ Rauch (Pasadena: Carnegie Observatories, 
	http://www.ociw.edu/ociw/symposia/series/symposium4/\\
	proceedings.html)

Fulbright, J.\ P.\ \& Johnson, J.\ A.\ 2003, ApJ, 595, 1154

Fusi Pecci, F., Bellazzini, M., Cacciari, C., \& Ferraro, F.\ R.\ 
	1995, AJ, 110 1664

Garz, T.\ 1973, A\&A, 26, 471

Geisler, D., Smith, V.\ V., Wallerstein, G., Gonzalez, G., \&
	Charbonnel, C.\ 2005, AJ, 129, 1428

Gonzalez, G.\ \& Wallerstein, G.\ 1998, AJ, 116, 765

Gratton, R.\ G.\ 1989, A\&A, 208, 171

Gratton, R.\ G., \etal\ 2001, A\&A, 369, 87

Gratton, R.\ G., Carretta, E., Eriksson, K., \& Gustafsson, B.\  1999,
A\&A, 350, 955 

Gratton, R.\ G.\ \& Sneden, C.\ 1991, A\&A, 241, 501

Gratton, R.\ G., Sneden, C., \& Carretta, E.\ 2004, \araa, 42, 385

Grevesse, N., Blackwell, D.\ E., \& Petford, A.\ D.\ 1989, A\&A, 208, 157

Grevesse, N., \& Sauval, A.\ J.\ 1998, Sp.\ Sci.\ Rev., 85, 161

Gustafsson, B., Edvardsson, B., Eriksson, K., Mizuno-Wiedner, M., 
	J\o rgensen, U. G., \& Plez, B. 2002, ASP Conf. Ser. Vol 288,
	ed. I. Hubeny,
	D. Mihalas, \& K. Werner (San Francisco: ASP) 331.

Hannaford, P., Lowe, R.\ M., Grevesse, N., Bi\'emont, E., \& Whaling, W.\ 
	1982, ApJ, 261, 736

Hanson, R.\ B., Sneden, C., Kraft, R.\ P., \& Fulbright, J.\ 1998, AJ,
	116, 1286 

Heise, C.\ \& Kock, M.\ 1990, A\&A, 230, 244

Hesser, J.\ E.\ 1978, ApJ, 223, L117

Hesser, J.\ E.\ \& Bell, R.\ A.\ 1980, ApJ, 238, L149

Hill, V., Francois, P., Spite, M., Primas, F., \& Spite, F.\ 2000, A\&A,
	364, 19

Hill, V., Pompeia, L., \& Spite, M.\ in press, in, ``Chemical abundances 
	and mixing in stars in the Milky Way and its satellites", ed.\ 
	L.\ Pasquini \& S.\ Randich, Springer Verlag Astrophysics Series


H\"oflich, P.\ A., Gerardy, C.\ L., Fesen, R.\ A.\ \& Sakai, S.\ 2002, 
	ApJ, 568, 791

H\"oflich, P.\ A., \& Khokhlov, A.\ 1996, ApJ, 457, 500

H\"oflich, P.\ A., Khokhlov, A.\ M.\ \& Wheeler, J.\ C.\ 1995, ApJ, 444, 831

H\"oflich, P.\ A., Wheeler, J.\ C.\ \& Thielemann, F.\ K.\ 1998, ApJ,
	495, 629

Houdashelt., M.\ L., Bell, R.\ A., \& Sweigart, A.\ V.\ 2000, AJ, 119, 1448

Ibata, R.\ A., Gilmore, G., \& Irwin, M.\ J.\ 1994, Nature, 370, 194

Ibata, R.\ A., Gilmore, G., \& Irwin, M.\ J.\ 1995, MNRAS, 277, 781

Ivans, I.\ I.\ in press, in, ``Chemical abundances and mixing in stars 
	in the Milky Way and its satellites", ed.\ L.\ Pasquini \& S.\ 
	Randich, Springer Verlag Astrophysics Series

Ivans, I.\ I., Kraft, R.\ P., Sneden, C., Smith, G.\ H., Rich, R.\ M., 
	\& Shetrone, M.\ 2001, AJ, 122, 1438 [M5-I01]

Ivans, I.\ I., Sneden, C., James, C.\ R., Preston, G.\ W., Fulbright,
	J.\ P., H\"oflich, P.\ A., Carney, B.\ W., \& Wheeler, J.\ C.\
	2003, ApJ, 592, 905

Ivans, I.\ I., Sneden, C., Kraft, R.\ P., Suntzeff, N.\ B., Smith, V.\ V., 
	Langer, G.\ E., \& Fulbright, J.\ P.\ 1999, AJ, 118, 1273 [M4-I99]

Iwamoto, K., Brachwitz, F., Nomoto, K., Kishimoto, N., Umeda, H., Hix, 
	W.\ R.\  \& Thielemann, F.-K.\ 1999, ApJS, 125, 439

James, G.\ \etal\ 2004, A\&A, 414, 107

Johnson, C.\ I., Kraft, R.\ P., Pilachowski, C.\ A., Sneden, C., Ivans,
I.\ I., \& Benman G.\ 2005, PASP, in press

Johnson, J.\ A.\ 2002, ApJS, 139, 219 

Johnson, J.\ A., Bolte, M., Stetson, P.\ B., Hesser, J.\ E., 
	Somerville, R.\ S.\ 1999, ApJ, 527, 199 [LMC-J99]

Kastberg, A., Villemoes, P., Arnesen, A., Heijkenskj\"old, F.,\&
	Langereis, A.\ 1993, J.\ Opt.\ Soc.\ AM.\ B, 10, 1330

Korn, A.\ 2004  in Carnegie Observatories Astrophysics Series, 
	Vol.\ 4: Origin and Evolution of the Elements, ed.\ A.\ 
	McWilliam and M.\ Rauch (Pasadena: Carnegie Observatories, 
	http://www.ociw.edu/ociw/symposia/series/symposium4/proceedings.html)

Kraft, R.\ P., \& Ivans, I.\ I.\ 2003, PASP, 115, 143 [Fe-KI03]

Kraft, R.\ P., \& Ivans, I.\ I.\ 2004, in Carnegie Observatories
	Astrophysics Series, Vol 4: Origin and Evolution of the
	Elements, ed.\ A.\ McWilliam and M.\ Rauch (Pasadena: Carnegie
	Observatories,
	http://www.ociw.edu/ociw/symposia/series/symposium4/proceedings.html)
	[Fe-KI04]

Kraft, R.\ P., Sneden, C., Langer, G.\ E., \& Shetrone, M.\ D.\ 
	1993, AJ, 106, 1490

Kraft, R.\ P., Sneden, C., Langer, G.\ E., \& Prosser, C.\ F.\ 1992,
	AJ, 104, 645

Kraft, R.\ P., Sneden, C., Langer, G.\ E., Shetrone, M.\ D.,
	\& Bolte, M.\ 1995, AJ, 109, 2586

Kraft, R.\ P., Sneden, C., Smith, G.\ H., Shetrone, M.\ D., 
	\& Fulbright, J.\ 1998, AJ, 115, 1500

Kroll, S., \& Kock, M.\ 1987, A\&AS, 67, 225


Kurucz, R.\ L.\ 1992, Rev.\ Mex.\ Astron.\ Astrofis., 23, 181

Kurucz, R.\ L.\ 1993, ATLAS9 Stellar Atmosphere Programs 
	and 2 km/s Grid, Kurucz CD-ROM \#13, Cambridge, MA: Smithsonian 
	Astrophysical Observatory

Kurucz, R.\ L., \& Bell, B.\ 1995, 1995 Atomic Line Data, Kurucz
	CD-ROM \#23, Cambridge, MA: Smithsonian Astrophysical Observatory

Langer, G.\ E., Hoffman, R.\ E., \& Zaidins, C.\ S.\ 1997, PASP, 109, 244

Lawler, J.\ E.\ \& Dakin, J.\ T.\ 1989, J.\ Opt.\ Soc.\ Am.\ B, 6, 1457

Lawler, J.\ E., Bonvallet, G., \& Sneden, C.\ 2001, ApJ, 556, 452

Lawler, J.\ E., Wickliffe, M.\ W., den Hartog, E.\ A., \& Sneden, C.\ 
	2001, ApJ, 563, 1075

Lef\'ebvre, P.-H., Garnir, H.-P., \& Bi\'emont, E.\ 2003, A\&A, 404, 1153

Lee, J.-W.\ \& Carney, B.\ W.\ 2002, AJ, 124, 1151

Lee, J.-W., Carney, B.\ W., \& Habgood, M.\ J.\ 2004, AJ, in press 
	(astro-ph/0409706)

Lin, D.\ N.\ C.\ \& Richer, H.\ B.\ 1992, ApJ, 388, L57

Lindblad, B.\ 1922, ApJ, 55, 85

Lloyd Evans, T.\ 1980, MNRAS,  193, 87

Mansour, N.\ B., Dinneen, T., Young, L., \& Cheng, K.\ T.\ 1989,
	Phy.\ Rev.\ A, 39, 5762

Martin, G.\ A., Fuhr, J.\ R., \& Wiese, W.\ L.\ 1988,
	J.\ Phys.\ Chem.\ Ref.\ Data, 17, 3

Mashonkina, L.\ I., Shimanski\u i, V.\ V., \& Sakhibullin, N.\ A.\ 
	2000, Astronomy Reports, 44, 790, trans.\ from 2000, 
	Astron.\ Zh., 77 893

Matteucci, F., Raiteri, C.\ M., Busso, M., Gallino, R., \& Gratton, R.\ 
	1993, A\&A, 272, 421

Matteucci, F.\ \& Recchi, S.\ 2001, \apj, 558, 351

May, M., Richter, J., \& Wichelmann, J.\ 1974, A\&AS, 18, 405

McWilliam, A.\ 1997, ARAA, 35, 503

McWilliam, A., Preston, G.\ W., Sneden, C., \& Searle, L.\ 1995, AJ, 109, 2757

McWilliam, A., Rich, R.\ M, \& Smecker-Hane, T.\ A.\ 2003, ApJ, 592, L21 

McWilliam, A.\ \& Smecker-Hane, T.\ A.\ 2005, \apj, 622, L29

Messenger, B.\ B., \& Lattanzio, J.\ C.\ 2002, \mnras, 331, 684

Mighell, K.\ J., Rich, R., M., Shara, M., \& Fall, S.\ M.\ 1996, AJ, 111, 2314

Mishenina, T.\ V., Kovtyukh, V.\ V., Soubiran, C., Travaglio, C., \&
	Busso, M.\ 2002, A\&A, 396, 189

Nakamura, T., Umeda, H., Iwamoto, K., Nomoto, K., Hashimoto, M., Hix,
	W.\ R., Thielemann, F.-K.\ 2001, ApJ, 555, 880

Nissen, P.\ E.\ \& Schuster W.\ J.\ 1997, A\&A, 326, 751 [NS97]

Nitz, D.\ E., Kunau, A.\ E., Wilson, K.\ L., \& Lentz, L.\ R.\ 1999, APJS,
	122, 557

Nitz, D.\ E., Wickliffe, M.\ E.\ \& Lawler, J.\ E.\ 1998, ApJS, 117, 313

O'Brian, T.\ R., Wickliffe, M.\ E., Lawler, J.\ E., Whaling, W., \&
	Brault, J.\ W.\ 1991, J.\ Opt.\ Soc.\ Am.\ B, 8, 1185

Olsen, K.\ A.\ G., Hodge, P.\ W., Mateo, M., Olszewski, E.\ W.,
	Schommer, R.\ A., Suntzeff, N.\ B., \& Walker, A.\ R.\ 1998,
	MNRAS, 300, 665 [LMC-O98]

Olszewski, E.\ W., Schommer, R.\ A., Suntzeff, N.B., \& Harris,
	H.\ C.\ 1991, AJ, 101, 515 [LMC-O91]

Osborn, W.\ 1971, Observatory, 91, 223

Palmeri, P., Bi\'emont, E., Abousa\"id, \& Godefroid, M.\ 1995,
	J.\ Phys.\ B, 28, 3741

Peterson, R.\ C.\ 1980, ApJ, 237, L87

Pickering, J.\ C.\ 1996, ApJS, 107, 811

Pickering, J.\ C., Thorne, A.\ P., \& Perez, R.\ 2001, ApJS, 132, 403

Pilachowski, C.\ A., Sneden, C., \& Wallerstein, G.\ 1983, ApJS, 52, 241

Popper, D.\ M.\ 1947, ApJ, 105, 204

Prochaska, J.\ X., Naumov, S.\ O., Carney, B.\ W., McWilliam, A., \&
	Wolfe, A.\ M.\ 2000, AJ, 120, 2513

Ram\'irez, S.\ V.\ \& Cohen, J.\ G.\ 2002, AJ, 123, 3277


Ryan, S.\ G., Norris, J.\ E.\ \& Beers, T.\ C.\ 1996, ApJ, 471, 254

Salpeter, E.\ E.\ 1955, ApJ, 121, 161

Sarajedini, A.\ 1994, AJ, 107, 618

Schnabel, R., Kock, M., \& Holweger, H.\ 1999, A\&A, 342, 610

Searle, L., Wilkinson, A., \& Bagnuolo, W.\ G.\ 1980, ApJ, 239, 803

Searle, L., \& Zinn, R.\ 1978, ApJ, 225, 357


Shetrone, M.\ D., C\^ot\'e, P., \& Sargent, W.\ L.\ W.\ 2001, ApJ, 548,
	592 [dSph-S01]


Shetrone, M.\ D.\ \& Keane, M.\ J.\ 2000, AJ, 119, 840


Shetrone, M.\ D., Venn, K.\ A., Tolstoy, E., Primas, F., Hill, V., \& 
	Kaufer, A.\ 2003, AJ, 125, 684

Simmerer, J., Sneden, C., Ivans, I.\ I., Kraft, R.\ P., Shetrone, M.\ D.,
	\& Smith, V.\ V.\ 2003, AJ, 125, 2018

Smecker-Hane, T.\ \& McWilliam, A.\ 2004, ApJ, submitted 

Smith, G.\ \& Raggett, D.\ St.\ J.\ 1981, J.\ Phys.\ B., 14, 4015

Smith, G.\ H., Sneden, C., \& Kraft, R.\ P.\ 2002, AJ, 123, 1502

Smith, V.\ \etal\ 2002, AJ, 124, 3241

Sneden, C.\ 1973, ApJ, 184, 839

Sneden, C.\ \& Crocker, D.\ A.\ 1988, ApJ, 335, 406

Sneden, C., Kraft, R.\ P., Guhathakurta, P., Peterson, R.\ C., \& Fulbright,
	J.\ P.\ 2004, AJ, 127, 2162

Sneden, C., Kraft, R.\ P., Shetrone, M.\ D., Smith, G> H., Langer, G.\ E.,
	\& Prosser, C.\ F.\  1997, AJ, 114, 1964

Stephens, A.\ 1999, AJ, 117, 1771

Stetson, P.\ B., Hesser, J.\ E., Smith, G.\ H., Vandenberg, D.\ A., \& 
	Bolte, M.\ 1989, AJ, 97, 1360

Sweigart, A.\ V.\ \& Mengel, J.\ G.\ 1979, ApJ, 229, 624

Takeda, Y., Zhao, G., Takada-Hidai, M., Chen, Y.-Q., Saito, Y., \& 
	Zhang, H.-W.\ 2003, Chin.\ J.\ Astron.\ Astrophys., 3, 316

Tautvai{\v s}ien{\.e}, G., Wallerstein, G., Geisler, D., Gonzalez, G., \& 
	Charbonnel, C.\ 2004, AJ, 127, 373 

Testa, V., Ferraro, F.\ R., Brocato, E., \& Castellani, V.\ 1995, MNRAS,
	275, 454

Timmes, F.\ X., Woosley, S.\ E., \& Weaver, T.\ A.\ 1195, ApJS, 998, 617

Tinsley, B.\ M.\ 1979, ApJ, 229, 1046

Tolstoy, E., Venn, K., Shetrone, M., Primas, F., Hill, V., Kaufer, A., \&
	Szeifert, T.\ 2003, AJ, 125, 707

Tozzi, G.\ P., Brunner, A.\ J., \& Huber, M.\ C.\ E.\ 1985, MNRAS, 217, 423

Truran, J.\ W.\ 1981, A\&A, 97, 391

van den Bergh, S.\ 1967, AJ, 72, 70

Whaling, W., Hannaford, P., Lowe, R.\ M., Bi\'emont, E., \& Grevesse,
	N.\ 1985, A\&A, 153, 109

Wickliffe, M.\ E.\ \& Lawler, J.\ E.\ 1997, ApJS, 110, 163

Woosley, S.\ E.\ \& Weaver, T.\ A.\ 1995, ApJS, 101, 181

\end{references}
\end{document}